\documentstyle[here,epsfig]{mn2e}
\def\simlt{\mathrel{\rlap{\lower 3pt\hbox{$\sim$}}\raise 2.0pt\hbox{$<$}}}
\def\simgt{\mathrel{\rlap{\lower 3pt\hbox{$\sim$}} \raise 2.0pt\hbox{$>$}}}

\def\Msun{M_{\odot}}

\def\gtsima{$\; \buildrel > \over \sim \;$}
\def\ltsima{$\; \buildrel < \over \sim \;$}
\def\gtrsim{\lower.5ex\hbox{\gtsima}}
\def\lesssim{\lower.5ex\hbox{\ltsima}}
\def\url#1{{\ttfamily\def\/{/\discretionary{}{}{}}#1}}
\def\etal{{\it et al.~}}

\begin{document}

\newcommand{\q}{\begin{equation}}
\newcommand{\qa}{\begin{eqnarray}}
\newcommand{\qs}{\begin{eqnarray*}}
\newcommand{\nq}{\end{equation}}
\newcommand{\nqa}{\end{eqnarray}}
\newcommand{\nqs}{\end{eqnarray*}}
\newcommand{\ud}{\mathrm{d}}

\title[Early BH radiation - I: effects on the IGM] 
{Radiation from early black holes - I: effects on the
  neutral inter-galactic medium}
\author[E. Ripamonti, M. Mapelli, S. Zaroubi]
{E. Ripamonti$^{1,2}$, M. Mapelli$^{3}$, S. Zaroubi$^{2}$\\
$^{1}$Dipartimento di Fisica, Universit\`a di Milano-Bicocca, Piazza
della Scienza 3, I-20123 Milano, Italy; {\tt ripa@astro.rug.nl}\\
$^{2}$Kapteyn Astronomical Institute, University of Groningen, Postbus
800, NL-9700AV, Groningen, The Netherlands\\
$^{3}$Institute for Theoretical Physics, University of Z\"urich,
Winterthurerstrasse 190, CH-8057 Z\"urich, Switzerland}

\maketitle \vspace {0.1truecm}

\begin{abstract}
In the pre-reionization Universe, the regions of the inter-galactic
medium (IGM) which are far from luminous sources are the last to undergo
reionization. Until then, they should be scarcely affected by stellar
radiation; instead, the X-ray emission from an early black hole (BH)
population can have much larger influence. We investigate the effects of
such emission, looking at a number of BH model populations (differing
for the cosmological density evolution of BHs, the BH properties, and
the spectral energy distribution of the BH emission). We find that BH
radiation can easily heat the IGM to $10^3-10^4\;{\rm K}$, while
achieving partial ionization. The most interesting consequence of this
heating is that BHs are expected to induce a 21-cm signal ($\delta
T_b\sim 20-30\;{\rm mK}$ at $z\lesssim12$) which should be observable
with forthcoming experiments (e.g. LOFAR). We also find that at
$z\lesssim 10$ BH emission strongly increases the critical mass
separating star-forming and non-star-forming halos.
\end{abstract}

\section{Introduction}\label{sec:introduction}

The Sloan Digital Sky Survey\footnote{\url http://www.sdss.org/} (SDSS)
has unveiled the existence of quasars at redshift $z\gtrsim{}6$ (Becker
et al. 2001; Djorgovski et al. 2001; Fan et al. 2001, 2002, 2003; White
et al. 2003).  This indicates that super-massive black holes (SMBHs)
with a mass of $10^{9-10}\,{}M_\odot{}$ had already formed when the
Universe was less than 1 Gyr old (Fan et al. 2001, 2003).

The processes which lead to the formation of such  huge black holes (BHs)
already in the early stages of the life of the Universe are very
uncertain. A possible scenario is that SMBHs were built up starting from
a seed intermediate-mass BH (IMBH, i.e. a BH with mass of
$20-10^5\,{}M_\odot{}$), which increased its mass by accreting gas
and/or by merging with other IMBHs.

In particular, if first stars are very massive ($>260\,{}M_\odot{}$)
their fate is to directly collapse into BHs, nearly without losing mass
(Heger \& Woosley 2002). This can produce a population of IMBHs, which
are expected to efficiently accrete gas in the high-density primordial
Universe and eventually to coalesce with other BHs (Volonteri, Haardt \&
Madau 2002, 2003; Islam, Taylor \& Silk 2003, 2004; Volonteri \& Perna
2005; Volonteri \& Rees 2005, 2006). Furthermore, the accretion of these
IMBHs might be enhanced also during galaxy mergers, which tend to drive
gas into the inner regions of the host galaxy (Madau et al. 2004).
However, recent simulations by Pelupessy, Di Matteo \& Ciardi (2007)
suggest that the accretion history of such seed IMBHs can hardly account
for the SMBHs of the SDSS.

On the other hand, seed BHs can be produced also by the direct collapse
of dense, low angular momentum gas (Haehnelt \& Rees 1993; Umemura, Loeb
\& Turner 1993; Loeb \& Rasio 1994; Eisenstein \& Loeb 1995; Bromm \&
Loeb 2003), driven by turbulence (Eisenstein \& Loeb 1995) or
gravitational instabilities (Koushiappas, Bullock \& Dekel 2004; 
Begelman, Volonteri \& Rees 2006, hereafter BVR06; Lodato \& Natarajan
2006). In particular, the so-called 'bars within bars' mechanism
(Shlosman, Frank \& Begelman 1989; Shlosman, Frank \& Begelman 1990)
implies that bars, which form in self-gravitating clouds under some
assumptions, can transport angular momentum outwards on a dynamical
time-scale via gravitational and hydrodynamical torques, allowing the
radius to shrink. This shrinking produces greater instability and the
process cascades. BVR06 show that
this process leads to the formation of a 'quasi-star', which rapidly
collapses into a $\sim{}20\,{}M_\odot{}$ BH at the center of the
halo. The BH should encounter very rapid growth due to efficient gas
accretion.  

This allows the formation of seed BHs with mass
$\lesssim{}10^6\,{}M_\odot{}$ (BVR06; Lodato \& Natarajan 2006, 2007),
depending on the initial parameters (e.g. the temperature of the gas and
the spin parameter of the parent halo).

If such seed BHs formed at high redshift ($z\sim{}10-30$), they likely
played a crucial role in the early Universe. Previous studies
(Machachek, Bryan \& Abel 2003; Madau et al. 2004; Ricotti \& Ostriker
2004, hereafter RO04; Ricotti, Ostriker \& Gnedin 2005, and references
therein) showed that IMBHs accreting as miniquasars could be important
sources of partial, early reionization. The efficiency of miniquasars in
reionizing the high redshift Universe is especially due to the hardness
of their spectra, which extend up to the X-ray band.

For this reason, miniquasars are also indicated as sources of the X-ray
background, and their density can be strongly constrained by the level
of the unresolved fraction of this background (RO04;
Dijkstra, Haiman \& Loeb 2004; Ricotti et al. 2005; Salvaterra, Haardt
\& Ferrara 2005; Volonteri, Salvaterra \& Haardt 2006; Salvaterra,
Haardt \& Volonteri 2007).

Finally, miniquasars can also heat the intergalactic medium (IGM; see
e.g.  Nusser 2005; Zaroubi et al. 2007, hereafter Z07+; Thomas \&
Zaroubi 2007), influencing a plethora of processes (the 21 cm line
emission/absorption, the formation of the first structures, etc.).

In this paper we analyze all the main effects that primordial
miniquasars can produce on the neutral IGM, i.e. on the regions of the 
Universe outside the ionized regions produced by the first BHs and stars. 
This is done
by means of semi-analytical models combined with hydro-dynamical
simulations. We consider all the most significant models for miniquasar
formation, density evolution and spectra. Whereas previous studies
mostly focused on single aspects, our aim is to give a global
description, as complete as possible, of the role played by IMBHs in the
early Universe.

In particular, in Section 2 we will present an estimate of the radiation
background produced by primordial BHs at high redshift.
Then, we will discuss its effects on the IGM evolution, and in
particular on the cosmic microwave background (CMB) spectra and on the 21 cm emission  (Section 3).
In Section 4 we will discuss whether the radiation background can delay
structure formation.  In Section 5 we will discuss the relevance of
our findings in the light of previous results.
Finally, our results will be summarized in Section 6.

We adopt the best-fitting cosmological parameters after the 3-yr WMAP
results (Spergel \etal 2007), i.e. $\Omega{}_{\rm b}=0.042$,
$\Omega{}_{\rm M}=0.24$, $\Omega{}_{\rm DM}\equiv{}\Omega{}_{\rm
M}-\Omega{}_{\rm b}=0.198$, $\Omega{}_\Lambda{}=0.76$, $h=0.73$,
$H_0=100\,{}h$ km s$^{-1}$ Mpc$^{-1}$.

\section{Energy injection into the neutral IGM}\label{sec:energy injection}

\subsection{The total energy input}\label{ssec:energy input}

First of all, we develop a formalism for estimating the total energy
input of BHs into the neutral IGM at a given redshift, starting from
basic properties of the BH population, taken from semi-analytic models
(see next Section), such as the BH mass density $\rho_{BH}$ at redshift
$z$, the average BH mass $\langle M_{BH} \rangle$ at redshift $z$, and
the duty cycle $y$ of single BHs.

We start from considering that the mean free path of a photon of
energy $E$ emitted at redshift $z$ is:
\begin{equation}
\lambda(E,z)=[n_B(z) \sigma(E)]^{-1},
\label{eq:mean free path}
\end{equation}
where $n_B(z)=n_B(0)(1+z)^3$ is the cosmological baryon number density
at redshift $z$ [$n_B(0)\simeq2.5\times10^{-7}\;{\rm cm^{-3}}$; Spergel
et al. 2007], and $\sigma(E)$ is the photo-ionization cross section per
baryon of the cosmological mixture of H and He, which is approximately
\begin{equation}
\sigma(E) =
\left\{{
  \begin{array}{ll}
    0.75 \sigma_H(E) &
    13.6\,{\rm eV}\le E \le 25\,{\rm eV}\\
    \sigma_{250}\,[E/(250\,{\rm eV})]^{-2.65} &
    25\,{\rm eV}\le E \le 250\,{\rm eV}\\
    \sigma_{250}\,[E/(250\,{\rm eV})]^{-3.3} &
    250\,{\rm eV}\le E,\\
\end{array}}\right .
\label{eq:cross section}
\end{equation}
where $\sigma_H(E)$ is the photo-ionization cross section of hydrogen
(see eq. 2.4 of Osterbrock 1989), and $\sigma_{250}\simeq
3.2\times10^{-21}\;{\rm cm^2}$ is the cross section for $250\;{\rm eV}$
photons (see Zdziarski \& Svensson 1989 for further details on the cross
section at $E>25\;{\rm eV}$). In this paper we will neglect the
absorption of photons with $E<13.6\;{\rm eV}$. It is important to note
that the above cross section is appropriate only for a neutral gas.
Since the IGM close to luminous sources is mostly ionized (and the
ionized fraction is not zero even in mostly neutral regions),
eq. (\ref{eq:mean free path}) might lead to an underestimation of
$\lambda$, but this effect is important only in the last phases of
reionization.

On the other hand, the average distance between `active' BHs is
\begin{equation}
d = \left[{{\rho_{BH}(z)\,y}\over
{{\mathcal C}\, \langle M_{BH} \rangle(z)}}\right]^{-1/3},
\label{eq:BHdistance}
\end{equation}
where ${\mathcal C}$ accounts for the clustering of BHs\footnote{If
primordial BHs form in clusters (typically) of $N_{cl}$ BHs, the
probability that at least one of them is in an active state is
$1-(1-y)^{N_{cl}}$ (rather than $y$); so ${\mathcal C}\simeq N_{cl}\, y
/ [1-(1-y)^{N_{cl}}]$. In the following we will use ${\mathcal C}=10$
for models where BH clustering should be strong; this corresponds to
$N_{cl}\simeq330(160)$ for $y=0.03(0.06)$.}.

The comparison of $\lambda$ and $d$ shows that for photons of
sufficiently high energy the mean free path can easily exceed $d$. For
instance, the mean free path of a $500\;{\rm eV}$ photon emitted at
$z=20$ is about 9 comoving Mpc ($\sim d$ in typical models), but at the
same redshift a $1\;{\rm keV}$ photon typically propagates for $\sim90$
comoving Mpc. Since BHs are believed to emit a significant fraction of
their luminosities at such energies, they will build up a roughly
uniform background radiation field.

The BH emissivity can be estimated by assuming that during active
phases of accretion, each primordial BH produces radiation at a
fraction $\eta$ of the Eddington luminosity
[$L_{Edd}\simeq1.3\times10^{38}\;{\rm erg\,s^{-1}}\,(M_{BH}/\Msun)$],
and that their average spectrum at redshift $z$ is described by some
function $F(E,z)$. Then, the proper emissivity is
\begin{equation}
j(E,z) = {\mathcal L}\, {{F(E,z)}\over{\int F(E',z) dE'}}\,
\rho_{BH,\odot}(z)\, \left({\eta\over{0.1}}\right)\, y\, (1+z)^3,
\label{eq:emissivity}
\end{equation}
where ${\mathcal L}\simeq 4.4\times10^{-37}\;{\rm erg\, s^{-1}\,
cm^{-3}}\,\Msun^{-1}$ is a normalization constant\footnote{${\mathcal
L}$ is chosen to be the emissivity (per ${\rm cm^3}$) when the BH
density is equal to $1\Msun$ per comoving ${\rm Mpc^3}$ and BHs are
assumed to accrete with efficiency 0.1. So, it is equal to $0.1\times
1.3\times10^{38} ({\rm erg\,s^{-1}} \Msun^{-1}) \times
[3.086\times10^{24} ({\rm cm/Mpc})]^{-3}$.}, and $\rho_{BH,\odot}(z)$ is
the BH density at redshift $z$, expressed in solar masses per cubic
comoving Mpc.

The mean specific intensity of the radiation background at the observed
energy $E$, as seen by an observer at redshift $z$, is then (cfr. eq.~2
of Haardt \& Madau 1996)
\begin{equation}
J(E,z)={1\over{4\pi}} \int_{z+\Delta z}^\infty dz'\, {{dl}\over{dz'}}\,
\left({{1+z}\over{1+z'}}\right)^3\,
j(E{{1+z'}\over{1+z}},z')\,
e^{-\tau},
\label{eq:background spectrum}
\end{equation}
where the cosmological proper line element at redshift $z'$ is
\begin{equation}
{{dl}\over{dz'}} = {c\over{H_0}}\, {1\over{1+z'}}\,
{1\over{[\Omega_M (1+z')^3 + \Omega_\Lambda]^{1/2}}},
\label{eq:cosmological line element}
\end{equation}
and $\tau=\tau(E,z,z')$ is the optical depth effectively crossed by a
photon emitted at redshift $z'$ and reaching redshift $z$ with an energy
$E$,
\begin{equation}
\tau\equiv\tau(E,z,z') = \int_z^{z'} d\tilde{z}\, {{dl}\over{d\tilde{z}}}\,
\sigma(E{{1+\tilde{z}}\over{1+z}})\, n_B(\tilde{z}).
\label{eq:tau}
\end{equation}

The definition of a cosmological background
would require that in eq.~(\ref{eq:background spectrum}) $\Delta z=0$;
but this is not appropriate for our purposes. In fact, as we already
mentioned, we will be looking at regions outside the ionized `bubbles'
produced by the first luminous sources. So, we will examine the effects
of the radiation background on baryons quite removed from any particular
source (we will refer to such baryons as the `neutral-IGM' baryons),
i.e. at a distance of the order of $d/2$. This is irrelevant for photons
with long mean free paths ($\lambda(E,z)\gtrsim d/2$), but is of
fundamental importance for photons with $\lambda\ll d/2$, which are
absorbed in the vicinity of the BHs. In short, we start the redshift
integration in eq.~(\ref{eq:background spectrum}) from $z+\Delta z$
(where $\Delta z$ is chosen so as to skip the integration over distances
$\le d/2$), rather than from $z$.

From the background spectrum $\phi(E,z)$ we can easily obtain the energy
input per baryon due to the absorption of background photons at redshift
$z$,
\begin{equation}
\epsilon(z) = 4\pi\, \int dE\, J(E,z)\, \sigma(E).
\label{eq:energy input}
\end{equation}
It must
be noted that our use of the cross section~(\ref{eq:cross section}) in
the estimates of $\tau$ (eq. \ref{eq:tau}) and of $\epsilon$
(eq. \ref{eq:energy input}) might induce two opposite errors. First of
all, $\tau$ is overestimated (and $J$ underestimated) when a significant
fraction of the cosmic volume is ionized. On the other hand, when the
absorbing medium is not completely neutral, we overestimate the fraction
of radiative energy which is actually intercepted by the baryons. The
former effect leads to a moderate underestimation of $\epsilon$,
starting at relatively high redshifts; the latter might lead to a large
overestimation of $\epsilon$, but only for models where the IGM ionized
fraction becomes quite large. We neglect both effects in our
calculations: our results will generally be mild underestimates of the
BH effects, except in the cases where the ionized fraction becomes large
(a condition where we will significantly overestimate the BH effects).

The energy input must be split into a fraction $f_{ion}$ going
into ionizations, a fraction $f_{heat}$ going into heating, and a
fraction $f_{exc}$ going into excitations. These fractions actually
depend on the energy $E$ of the absorbed photon; but Shull \& van
Steenberg (1985) determined that, for all $E\gtrsim 100\;{\rm eV}$, they
are reasonably fitted by the expressions\footnote{We report the
expressions which are given for H, and neglect the small correction due
to the presence of He.}
\begin{eqnarray}
f_{heat} & = & 0.9971 [1 - (1-x_H^{0.2663})^{1.3163}]\\
f_{ion}  & = & 0.3908 (1-x_H^{0.4092})^{1.7592}\\
f_{exc}  & = & 0.4766 (1-x_H^{0.2735})^{1.5221}.
\label{eq:input partition}
\end{eqnarray}
where $x_H$ is the hydrogen ionization fraction
($x_H=n(H^+)/[n(H^0)+n(H^+)]$).  As can be seen in Figs.~\ref{fig:fig1}
and~\ref{fig:fig2}, the contribution of photons with $E<100\;{\rm eV}$
to the background is small or negligible (in the absence of
reionization, the mean free path of $100\;{\rm eV}$ photons exceeds
$\sim 1$ comoving Mpc and becomes comparable with $d$ only at $z\lesssim
5$): so the use of these energy-independent functions is legitimate.

\subsection{Model parameters}\label{ssec:model parameters}

\begin{table}
\begin{center}
\caption{Parameters of the BH growth model histories.}
\begin{tabular}{lccccc}
\hline
\vspace{0.1cm}
Model & $\rho_{BH}$$^a$ & $\langle M_{BH}\rangle$$^b$ & $y$$^c$ &
${\mathcal C}$ & $z_{s}$\\
\hline
IMBH-3\%  & $10^{6.75-0.275z}$      & $10^{3.5-0.05z}$ & 0.03 & 10 & 30\\
IMBH-6\%  & $10^{7.5-0.3z}$   & $10^{5-0.1z}$ & 0.06 & 10 & 30\\
SMBH-3\%  & $10^{7.25-0.375z}$   & $10^{6.5-0.05z}$ & 0.03 & 1  & 30\\
BVR06$^d$ & $10^{5.675-0.1875z}$ & $10^6$           & 0.10 & 1  & 18\\
\hline
\end{tabular}
\label{tab:tab1}
\end{center}
$^a$ In solar masses per cubic comoving Mpc.\\
$^b$ In solar masses.\\
$^c$ Consistent with the assumptions of the underlying models.\\
$^d$ Fig. 2 of BVR06 does not show the $\rho_{BH}$ evolution for $z<10$;
for this reason, at such redshift we will extrapolate the above formula.
\end{table}

In the above section we have seen how we can obtain an estimate of the
cosmological X-ray background produced by primordial BHs, and of the
energy it can inject in the IGM. Such estimate mainly depends on three
input quantities: the evolution of the cosmological density of BHs
$\rho_{BH}(z)$, the duty cycle $y$, and the typical spectral shape of an
active BH, $F(E)$. The evolution of the BH average mass $\langle
M_{BH}\rangle(z)$, and the clustering factor ${\mathcal C}$ have much
smaller effects.

\subsubsection{BH growth history}

There exist several models (e.g.  RO04; BVR06; Z07+) predicting the
evolution of the BH mass density in the early Universe. Here we will
discuss three different histories which are reasonable approximations of
the models IMBH-3\%, IMBH-6\%, and SMBH-3\% discussed in Z07+ (see their
fig. 8), and of one of the models in BVR06 (duty cycle 0.1, Mestel disk;
from their fig. 2). The two IMBH models (IMBH-3\% and IMBH-6\%) assume
that primordial BHs with mass $\sim 100\Msun$ form in small
($10^6-10^7\Msun$) and numerous halos, where H$_2$ cooling is efficient;
the SMBH-3\% and the BVR06 models, instead, assume that primordial BHs
of large mass ($\gtrsim 10^5\Msun$) form in larger ($10^8-10^9\Msun$)
and rarer halos cooled by atomic H. In all the four cases we will adopt
the simple power-law approximations of the Z07+ and BVR06 results which
are given in Table~\ref{tab:tab1}.  Such power-laws provide good fits to
all the original models for $z\ge10$, whereas at lower redshifts they
are either a reasonable extrapolation (for BVR06), or give a
slight-to-moderate underestimate of the predictions of the Z07+ models.
Table~\ref{tab:tab1} lists also the other parameters defining the BH
growth histories: the duty cycle $y$ is by far the most important,
whereas our results are relatively insensitive to the assumptions on
$\langle M_{BH}\rangle(z)$ and ${\mathcal C}$. This is quite fortunate
as the value of $y$ is intrinsic in our reference models, whereas none
of them provides a simple estimate of the other parameters. The values
given in Table~\ref{tab:tab1} are guesses based on the general
properties of the reference models, and on the notion that the typical
BH mass should increase with time (especially when $\rho_{BH}$ grows
fast).

Furthermore, we assume that before a certain redshift $z_{s}$
($z_{s}=30$ for the evolutions taken from Z07+, $z_{s}=18$ for
the one from BVR06) the BH density (and emissivity) is 0.

\subsubsection{BH spectral energy distribution}
We experiment with three different types of BH spectral energy
distributions (SEDs): simple power-laws $F(E,z)=F_\alpha(E)$, a template
$F(E,z)=F_{SOS,\alpha}(E)$ introduced by Sazonov, Ostriker \& Sunyaev
(2004), and a multi-component spectrum which is the sum of a multi-color
black body and a power-law spectrum (see Shakura \& Sunyaev 1973, and
Salvaterra et al. 2005), $F(E,z)=F_{MC,\Phi}(E,z)$.

Power-laws are characterized by their slope $\alpha$, and are assumed
to be
\begin{equation}
F_\alpha(E) =
\left\{{
  \begin{array}{ll}
    E^{-\alpha} & 0.01\;{\rm eV}<E<10^6\;{\rm eV}\\
    0 & {\rm otherwise.}\\
\end{array}}\right .
\end{equation}
In the following, we will consider the power-law with $\alpha=1$
(hereafter PL1 SED) as our reference spectrum.

The spectral template by Sazonov et al. (2004) is
characterized by the slope in the $1-100\;{\rm keV}$ range, and its
exact shape is
\begin{equation}
F_{SOS,\alpha}(E) =
\left\{{
  \begin{array}{ll}
    C_0             & 0.1\;{\rm eV}<E\le 10\;{\rm eV}\\
    E^{-1.7}        & 10\;{\rm eV}<E\le 10^3\;{\rm eV}\\
    C_1 E^{-\alpha} & 10^3\;{\rm eV}<E\le 10^5\;{\rm eV}\\
    C_2 E^{-1.6}    & 10^5\;{\rm eV}<E<10^6\;{\rm eV}\\
    0                     & {\rm otherwise},
\end{array}}\right .
\end{equation}
where the constants $C_1=10^{3(\alpha-1.7)}$ and $C_2=10^{2.9-2\alpha}$
are chosen so as to ensure continuity, and the constant
$C_0\simeq0.1607$ ensures that the fraction of the BH luminosity which
goes into photons with $E\le 10\;{\rm eV}$ is the same as in the
complete Sazonov et al. (2004) template (i.e. about 0.85), even though we
are not interested in the details of their model for $E\le 10\;{\rm
eV}$. In this paper we considered the case $\alpha=1$ (SOS1 SED): we
chose such a relatively steep value (Sazonov et al. 2004 suggest values
of about $0.7-0.8$) because such SED is intended to show what happens
with the steepest spectrum reasonably expected from BHs. However, our
results depend only weakly on this index.

Finally, in the multi-component SED, the multi-color black body
component is $\propto E^{1/3}$ and dominates up to a peak energy
\begin{equation}
E_p\simeq 3000\;{\rm eV} (M_{BH}/\Msun)^{-1/4}.
\label{eq:Epeak}
\end{equation}
Above that the multi-color black body is exponentially cutoff, and the
power-law component ($\propto E^{-1}$) emerges, as described in Shakura
\& Sunyaev 1973 (see also Salvaterra et al. 2005 for an
application to a context similar to the one we are considering):
\begin{equation}
F_{MC,\Phi}(E,z) = 
\left\{{
  \begin{array}{ll}
    E^{1\over3} e^{-{E\over{3 E_p}}} & 0.01\;{\rm eV}<E\le E_p\\
    E^{1\over3} e^{-{E\over{3 E_p}}} + A_\Phi E^{-1} & E_p<E\le 10^6\;{\rm eV}\\
    0 & {\rm otherwise.}
\end{array}}\right .
\end{equation}
$A_\Phi$ is chosen so that the energy in the power-law spectral
component is equal to a fraction $\Phi$ of the energy in the multi-color
black body spectral component. $\Phi$ is usually taken to be
$\lesssim{}1$, and we will consider the case $\Phi=0.1$ (MC01 SED),
which is practically indistinguishable from all the cases with lower
$\Phi$, and quite similar to the case with $\Phi=1$, too. As we
substitute $M_{BH}$ with $\langle M_{BH}\rangle(z)$ inside
eq.~(\ref{eq:Epeak}), we note that this spectral shape is slightly
dependent on redshift\footnote{The low-energy tail of a modified
blackbody spectrum is expected to be $\propto E^{2}$. We neglect such
slope change, as only a small fraction of the BH luminosity is emitted
in this region of the spectrum. We also note that the exponential
constant was chosen to be $3E_p$ in order to ensure that the multi-color
black body component actually has a (broad) peak at $E=E_p$, as
described in Salvaterra et al. (2005).}.

\begin{figure}
\vspace{-0.5truecm}
\center{{
\epsfig{figure=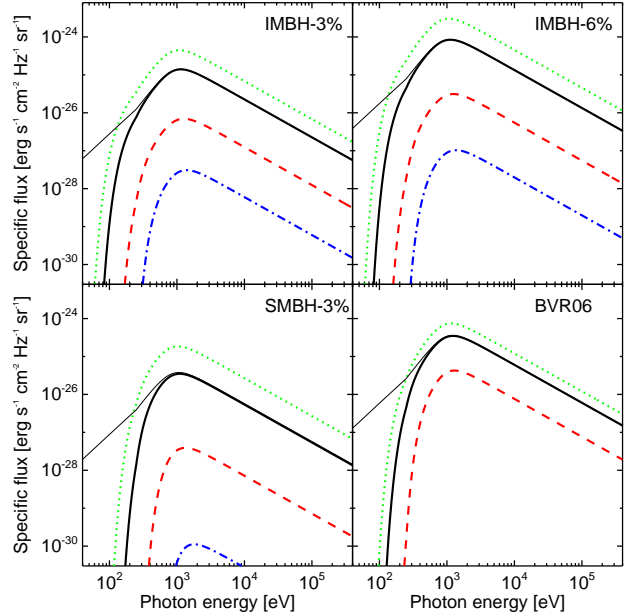,height=8.7cm}
}}
\caption{\label{fig:fig1} Spectrum of the background produced by
primordial accreting BHs and seen by a neutral-IGM baryon in
the four BH growth history scenarios (IMBH-3\%: top left panel;
IMBH-6\%: top right panel; SMBH-3\%: bottom left panel; BVR06: bottom
right panel) at various redshifts ($z=8$: thick dotted line; $z=10$:
thick solid line; $z=15$: thick dashed line; $z=20$: thick dot-dashed
line), assuming a PL1 spectrum for the BH emission. The thin solid line
shows the spectrum we would obtain at $z=10$, had we assumed that $\Delta
z=0$ in eq.~(\ref{eq:background spectrum}).}
\end{figure}

For all the considered spectral shapes we chose to assume that the BH
emissivity at energies below $0.01\;{\rm eV}$ or above $10^6\;{\rm eV}$
is 0. Such choice prevents numerical problems, and does not
significantly affect our results.

\subsection{Results}

\begin{figure}
\center{{
\vspace{-0.5truecm}
\epsfig{figure=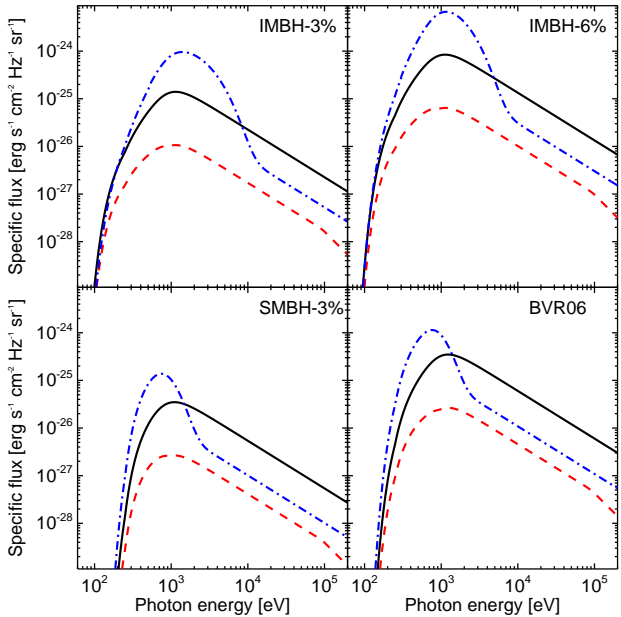,height=8.7cm}
}}
\caption{\label{fig:fig2} Spectrum of the background produced by
primordial accreting BHs, and seen by neutral-IGM baryons at
$z=10$. The four panels refer to the four BH growth history scenarios we
consider (IMBH-3\%: top left panel;IMBH-6\%: top right panel; SMBH-3\%:
bottom left panel; BVR06: bottom right panel), whereas the different
line types refer to three different SED for the BH spectra (PL1: solid
line; SOS1: dashed line; MC01: dot-dashed line).}
\end{figure}

In Fig.~\ref{fig:fig1} we show the redshift evolution of the spectrum of
the background radiation produced by primordial BHs and reaching a
neutral-IGM baryon. In such Figure we consider all the different BH
growth histories, but only the PL1 SED. The background level grows with
time, as could be expected when we remember that the BH density, the
average distance among BHs, and the IGM density all evolve in a
background-enhancing direction. In all the considered growth
scenarios, the spectra peak at $E\sim1\;{\rm keV}$ at all redshifts:
above this peak the spectrum is almost un-absorbed (i.e., the
specific flux decreases in the same way as the input spectrum), whereas
the spectrum at energies below the peak is shaped by the cutoff due to
the IGM absorption. Such absorption cutoff slowly moves to lower
energies. It is also interesting to look at the thin solid line, which
illustrates the effect of using $\Delta z=0$ in eq.~(\ref{eq:background
spectrum}): as expected, the high energy part of the spectrum does not
change, while the sharp cutoff at low energies is replaced by a much
milder power-law decline.

When comparing different BH growth histories in Fig.~\ref{fig:fig1}, it
is clear that the normalization of the background spectrum is related to
the value of the product $(y \times \rho_{BH})$ at the relevant
$z$. Instead, the sharpness of the low-energy cutoff depends on the
geometrical properties of the BH spatial distribution: in models with
large values of $\langle M_{BH}\rangle$ the cutoff is very sharp,
whereas it is a bit more gentle for IMBH models with low $\langle
M_{BH}\rangle$. This is important because the low energy part of the
spectrum, albeit accounting only for a small fraction of the total
energy in the background, is absorbed with quite high efficiency and is
a major contributor to the energy input $\epsilon$.

In Fig.~\ref{fig:fig2} we show the spectrum of the background radiation
at a fixed redshift, $z=10$, while varying the BH SED. It is clear that
`flat' (PL1, MC01) SEDs produce larger backgrounds than `steep'
(SOS1) SEDs, simply because a larger fraction of their luminosity is
emitted in the energy range ($E\gtrsim100-1000\;{\rm eV}$) where
incomplete (or negligible) absorption allows the build-up of the
background. This is particularly clear for the multi-component SEDs,
that produce a quite prominent bump in a broad energy range around
the $1\;{\rm keV}$ peak of the spectrum. Even if the peak of the
background specific flux is never far from $\sim 1\;{\rm keV}$, it is
possible to discern some trends: the SOS1 SED tends to peak at slightly
lower energies than the PL1 SED, whereas the position of peak of the
MC01 SED depends on the chosen BH growth scenario, simply because its
peak energy $E_p$ depends on the typical BH mass.

\begin{figure}
\center{{
\vspace{-0.5truecm}
\epsfig{figure=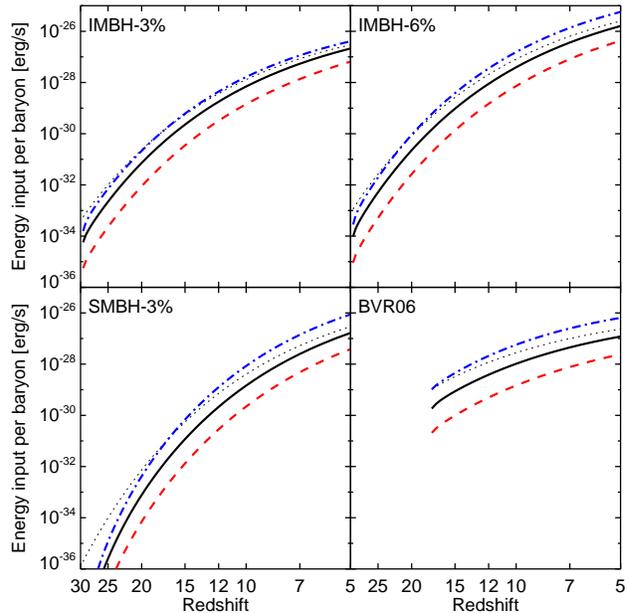,height=8.7cm}
}}
\caption{\label{fig:fig3} Redshift evolution of the total energy input
per neutral-IGM baryon due to the background produced by primordial BHs.
As explained in the caption of Fig.~\ref{fig:fig2}, the four panels
refer to the four considered BH growth histories, and each line type
refers to a different assumed BH spectrum. We also show the energy input
from BHs with a PL1 SED, had we assumed that $\Delta z=0$ in
eq.~(\ref{eq:background spectrum}) (thin dotted line).}
\end{figure}

\begin{table*}
\begin{center}
\caption{Fraction of the unresolved X-ray background which in our models
is due to BH emission at $z\ge z_{drop}$, in various bands. The numbers
in parenthesis are lower limits, obtained from the assumption that the
actual X-ray background is $1-\sigma$ higher than the central
values. The cases where the emission from our models exceeds the
unresolved background are in bold face.}
\begin{tabular}{lcccccc}
\hline
\vspace{0.1cm}
Model & $z_{drop}$ & 0.5-2 keV$^a$ & 2-8 keV$^b$ & 1-2 keV$^c$ & 2-5 keV$^d$ & 0.65-1 keV$^e$\\
\hline
IMBH-3\%+PL1  &5& 0.32(0.21)   & 0.19(0.10)   & 0.37(0.26)   & 0.42(0.13)   & 0.078(0.065)\\
IMBH-3\%+SOS1 &5& 0.024(0.016) & 0.014(0.008) & 0.028(0.020) & 0.031(0.010) & 0.006(0.005)\\
IMBH-3\%+MC01 &5& 0.67(0.44)   & 0.043(0.023) & 0.24(0.17)   & 0.096(0.030) & 0.21(0.18)\\
IMBH-3\%+MC01 &6& 0.18(0.11)   & 0.014(0.008) & 0.057(0.041) & 0.031(0.010) & 0.054(0.045)\\
IMBH-3\%+MC01 &7& 0.046(0.033) & 0.005(0.003) & 0.016(0.011) & 0.011(0.003) & 0.015(0.012)\\
\\
IMBH-6\%+PL1  &5&{\bf2.5(1.7)} &{\bf1.5(0.82)}&{\bf2.9(2.1)} &{\bf3.3(1.0)} & 0.63(0.52)\\
IMBH-6\%+PL1  &6& 0.79(0.51)   & 0.47(0.25)   & 0.92(0.65)   &{\bf1.0(0.32)}& 0.19(0.16)\\
IMBH-6\%+PL1  &7& 0.26(0.17)   & 0.16(0.084)  & 0.30(0.21)   & 0.34(0.10)   & 0.064(0.053)\\
IMBH-6\%+SOS1 &5& 0.19(0.13)   & 0.12(0.062)  & 0.22(0.16)   & 0.25(0.078)  & 0.048(0.040)\\
IMBH-6\%+MC01 &5&{\bf1.1(0.71)}& 0.33(0.18)   & 0.65(0.46)   & 0.72(0.22)   & 0.24(0.20)\\
IMBH-6\%+MC01 &6& 0.29(0.19)   & 0.10(0.055)  & 0.20(0.14)   & 0.23(0.069)  & 0.063(0.052)\\
IMBH-6\%+MC01 &7& 0.085(0.055) & 0.034(0.018) & 0.067(0.062) & 0.075(0.023) & 0.018(0.015)\\
\\
SMBH-3\%+PL1  &5& 0.26(0.17)   & 0.16(0.085)  & 0.30(0.22)   & 0.34(0.11)   & 0.065(0.054)\\
SMBH-3\%+SOS1 &5& 0.020(0.13)  & 0.012(0.006) & 0.023(0.016) & 0.026(0.008) & 0.005(0.004)\\
SMBH-3\%+MC01 &5& 0.048(0.031) & 0.029(0.016) & 0.055(0.039) & 0.063(0.019) & 0.012(0.010)\\
\\
BVR06+PL1     &5& 0.30(0.19)   & 0.18(0.097)  & 0.35(0.25)   & 0.39(0.12)   & 0.074(0.061)\\
BVR06+SOS1    &5& 0.023(0.015) & 0.014(0.007) & 0.027(0.019) & 0.030(0.009) & 0.005(0.004)\\
BVR06+MC01    &5& 0.054(0.035) & 0.032(0.018) & 0.063(0.044) & 0.071(0.022) & 0.013(0.011)\\
\hline
\end{tabular}
\label{tab:tab3}
\end{center}
$^a$ Flux in the 0.5-2 keV band, normalized to a background level of
$2.59(4.00)\times10^{-9}\;{\rm erg\,s^{-1}\,cm^{-2}\,sr^{-1}}$ (from the
combination of the B04 unresolved fraction, and the
Moretti et al. 2003 total X-ray background).\\
$^b$ Flux in the 2-8 keV band, normalized to a background level of
$4.35(8.05)\times10^{-9}\;{\rm erg\,s^{-1}\,cm^{-2}\,sr^{-1}}$, (from
the combination of the B04 unresolved fraction, and the De
Luca \& Molendi 2004 total X-ray background).\\
$^c$ Flux in the 1-2 keV band, normalized to a background level of
$1.12(1.58)\times10^{-9}\;{\rm erg\,s^{-1}\,cm^{-2}\,sr^{-1}}$ (from
HM07).\\
$^d$ Flux in the 2-5 keV band, normalized to a background level of
$1.31(4.26)\times10^{-9}\;{\rm erg\,s^{-1}\,cm^{-2}\,sr^{-1}}$ (from
HM07).\\
$^e$ Flux in the 0.65-1 keV band, normalized to a background level of
$3.28(3.94)\times10^{-9}\;{\rm erg\,s^{-1}\,cm^{-2}\,sr^{-1}}$ (from
HM07).\\
\end{table*}

Finally, in Fig.~\ref{fig:fig3} we show the total energy input per
baryon as a function of redshift. Such a quantity is well correlated
with the intensity of the background spectrum, especially at low
energies. Thus, it increases with time, and the SED with the highest low
energy component (MC01) gives the maximum energy input. We also compare
the energy input from the reference PL1 SED with that from an otherwise
identical model where we assumed that $\Delta z=0$ in
eq.~(\ref{eq:background spectrum}). This is useful to check the effects
of our assumption about $\Delta z$, and also gives us a rough estimate
of the level of the spatial fluctuations of the energy input. The
difference usually amounts to a factor of $2-3$, even if it might be larger
for the SMBH-3\% and the BVR06 growth histories, especially at high $z$.

We note that the model where the energy input from BHs is maximum is the
one where the IMBH-6\% accretion history is combined with the MC01 SED.
In the following, we will refer to such combination as the `extreme'
model, since it leads to the strongest BH feedback effects (and is also
close to the constraints from the unresolved X-ray background; see
below). On the other hand, we will also consider the IMBH-3\%+PL1 model
(i.e. the one combining an IMBH-3\% history with a PL1 SED) as a
`fiducial' case.

\subsubsection{Constraints from the unresolved X-ray background}

As a consistency check, we looked at whether our models are compatible
with measurements of the unresolved X-ray background from Bauer et
al. (2004; hereafter B04), and from Hickox \& Markevitch (2007;
hereafter HM07).

For such comparison, we obtained the spectrum of the background produced
by the BH emission at a redshift $z_{drop}$, we integrated it in the
relevant energy band, and we redshifted it to $z=0$ assuming no
absorption.

Such a calculation implies that, at $z\le z_{drop}$, the emissivity due
to BHs is 0. This is a quite crude assumption, but it must be remarked
that observations (Steidel et al. 2002) suggest that the duty cycle
declines with redshift (reaching $y\sim10^{-3}$ at $z=0$), and that
several theoretical models include a variation of $y$ (e.g., in model M3
of RO04 $y=1$ at $z\gtrsim14$, but $y=10^{-3}$ at $z\lesssim 7$).  It is
also possible that at redshifts $\lesssim 5-7$ an increasing
fraction of the BH sources are detected as resolved AGN sources. We also
stress that a fraction of IMBHs are expected to merge into larger SMBHs
or to be ejected from the parent halos as a consequence of three-body
encounters (see e.g. Volonteri et al. 2002, 2003). In these cases the
IMBHs no longer contribute to the X-ray background. As our model does
not account for these effects, it represents a strong upper limit for
the X-ray background from IMBHs.

The results of the comparison are listed in Table~\ref{tab:tab3}, where
we generally adopted $z_{drop}=5$. It can be seen that two of our models
(IMBH-6\% growth history, combined with either a PL1 or a MC01 SED)
exceed the observed background in at least one band. For such cases (and
also for the IMBH-3\%+MC01 case, where the contribution from BHs exceeds
half of the unresolved X-ray background in the $0.5-2\,{\rm keV}$ band)
we also list the result we would obtain with $z_{drop}=6$ or $7$, which
clearly show that the constraints from the X-ray background can be
easily satisfied also by these models, provided that $z_{drop}\gtrsim
6$. Thus, we note that the choice of $z_{drop}$ (and, in general, the
fate of IMBHs in the lower redshift range we consider) is quite
crucial for our models.  In the rest of this paper we will use
$z_{drop}=5$ for all the models, and our plots will extend to such
redshift.

\section{Influence on the IGM evolution}

\begin{figure}
\center{{
\vspace{-0.5truecm}
\epsfig{figure=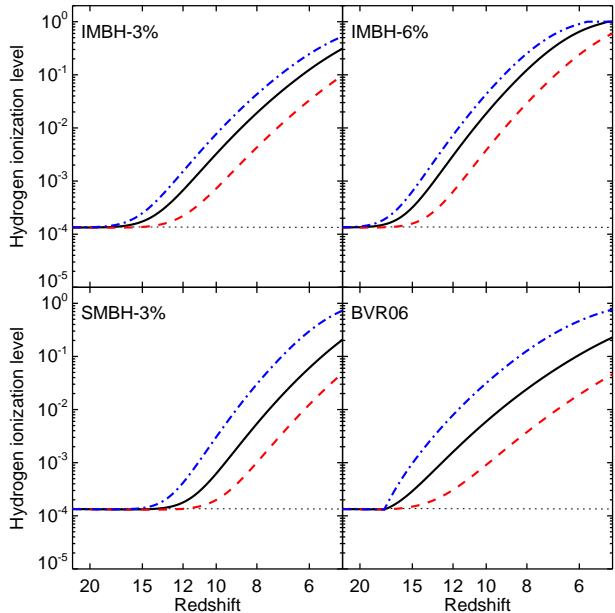,height=8.7cm}
}}
\caption{Redshift evolution of the hydrogen ionization fraction
$x_H$. The order of the panels and the meaning of the various line types
are the same as in Fig. \ref{fig:fig2}, except for the thin dotted
line, which represents the ionization evolution in a model without any
BH emission.}
\label{fig:fig4} 
\end{figure}

We looked at the effects of the energy input due to the background
radiation produced by primordial BHs on the thermal and chemical
evolution of the IGM. We employed a simplified version of the code
described in Ripamonti, Mapelli \& Ferrara (2007; hereafter RMF07; but
see Ripamonti et al. 2002, and Ripamonti 2007 - hereafter R07 - for more
detailed description of this code), in order to look at the evolution
of the IGM under the influence of the energy input we calculated in
the previous Section.

Such a code follows the gas thermal and chemical evolution.  The
chemistry part deals with 12 chemical species (H$^0$, H$^+$, H$^-$,
D$^0$, D$^+$, He$^0$, He$^+$, He$^{++}$, H$_2$, H$_2^+$, HD, and
e$^-$), and includes all of the reactions involving these species which
are listed in the Galli \& Palla (1998) minimal model for the primordial
gas, plus some important extension (e.g., it considers the ionizations
and the dissociations due to the energy input we are introducing).  The
thermal part includes the cooling (or heating, if the matter temperature
is lower than the CMB temperature) due to molecules (H$_2$ and HD), to
the emission from H and He atoms, to the scattering of CMB photons off
free electrons, and to bremsstrahlung radiation. Furthermore, it
accounts for the cooling/heating due to chemical reactions, and for the
heating due to the energy input we are considering\footnote{Other than
introducing the energy input as calculated in the previous Section, the 
code differs from the version described in RMF07 because we introduced
the cooling through He lines and bremsstrahlung (the rates were taken
from Anninos et al. 1997), and we splitted the energy input into the
heating, ionization and excitation components by using the expressions
given in eq.~(\ref{eq:input partition}), rather than the Chen \&
Kamionkowski (2004) approximations.}.

\begin{figure}
\center{{
\vspace{-0.5truecm}
\epsfig{figure=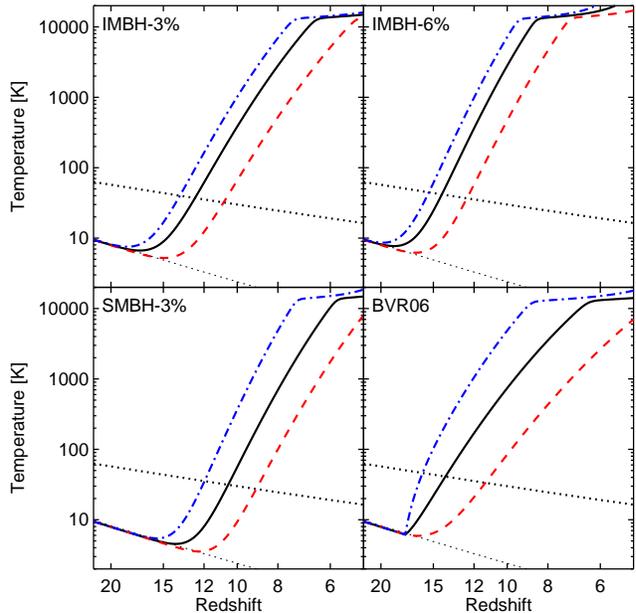,height=8.7cm}
}}
\caption{Redshift evolution of the IGM temperature $T_k$ in regions
outside ionized bubbles. The order of the panels and the meaning of the
various line types are the same as in Fig. \ref{fig:fig2}, except for
the thin dotted line, which represents the temperature evolution in a
model without any BH emission, and for the thick dotted line
representing the CMB temperature.}
\label{fig:fig5} 
\end{figure}

\subsection{IGM ionization and temperature}

Figs.~\ref{fig:fig4} and \ref{fig:fig5} show the effects of the BH
emission upon the ionization level (in particular, the hydrogen ionized
fraction $x_H=n(H^+)/[n(H^0)+n(H^+)]$ ) and the temperature of the IGM
$T_k$ (all these quantities are calculated outside the ionized bubbles
close to radiation sources). In all the models we consider, the BH
emission starts altering the neutral IGM at $z\sim15-20$. After that
there is a steady increase in both $x_H$ and $T_k$. The increase of
$x_H$ stops only when the IGM is completely ionized (however, such
condition is reached only in the most extreme of our models, and only at
a redshift $\sim 6$). Instead, $T_k$ stops increasing once it reaches a
level ($\sim10^4\,{\rm K}$), where atomic cooling is important: in the
models where BH emission is assumed to be strongest (IMBH-6\% with MC01
spectrum) this happens at $z\sim 10$, but $z\sim6-8$ is a more typical
range.

It must be noted that in the lower redshift range we consider (say
$z\lesssim10$) our models start suffering from several problems. First
of all, the energy input we employ is calculated for a neutral medium,
whereas in some of our models $x_H\gtrsim0.5$ already at
$z\sim7-8$. Then, we are overestimating the energy input\footnote{The
situation is actually quite complicated. The reduction in the heating
rate is slower than what could naively be expected [$\epsilon\propto
(1-x_H)^{-1}$] from the increase of $x_H$, because the bulk of the cross
section is due to He, which is harder to strip of its electrons (see
e.g. Thomas \& Zaroubi 2007). For example, in the `extreme' model, at
redshift 6 $x_H\simeq0.9$, but $x_{He++}\equiv
n(He^{++})/[n(He^0)+n(He^+)+n(He^{++})]\simeq 0.1$, and by using
eq.~(\ref{eq:cross section}) we are overestimating the energy input
$\epsilon$ only by a factor of $\sim2.5$ rather than $(1-x_H)^{-1}\sim10$.
Furthermore, the assumption that the IGM is completely neutral also
leads to an overestimation of the optical depth $\tau$, and the
background radiation (and energy input) is correspondingly
underestimated.  Then, our energy input rates are essentially correct,
except for our `extreme' model at $z\lesssim{}7$, where we might be
overestimating $\epsilon$ by a factor of $\lesssim 2$.}. Second, we are
assuming that the IGM density remains constant at its average
unperturbed cosmological value, whereas this approximation becomes
increasingly problematic as structures start to form. Third, there might
be some level of metal enrichment (altering both the heating and the
cooling rates) even in regions which are far away from the most luminous
sources. Finally, we are completely neglecting the contribution to
heating and reionization which is due to stars, which is likely to be
substantial at  relatively low redshifts. However, our calculations
should still be reasonably accurate until the end of the so-called
'overlap' phase of reionization (probably not far from
$z\sim{}7-8$, see Section 5.1),
provided that they are taken to represent conditions in regions which
were not yet ionized.

\subsection{CMB angular spectrum}

\begin{figure}
\center{{
\vspace{-0.5truecm}
\epsfig{figure=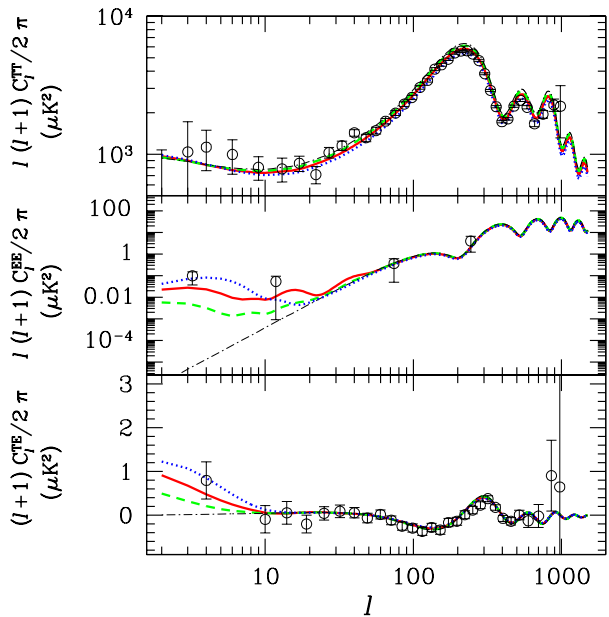,height=8.7cm}
}}
\caption{Effects of BH emission upon the CMB angular
spectra. Temperature-temperature (top panel), polarization-polarization
(central panel) and temperature-polarization (bottom panel) spectra are
shown. Thick dotted line: CMB spectra derived assuming Thomson optical
depth $\tau{}_e=0.09$ and a sudden reionization model (consistent with
the 3-yr {\it WMAP} data); thick solid and dashed lines: CMB
spectra derived assuming energy injection from the BHs in the
`extreme' (IMBH-6\%+MC01, i.e. the case where BH energy input is
strongest) and the `fiducial' (IMBH-3\%+PL1) cases, respectively; thin
dash-dotted line: CMB spectra derived assuming no reionization and no
contribution from BHs.}
\label{fig:fig6} 
\end{figure}

The cosmic heating and the contribution to reionization due to BHs might also leave some imprint on the CMB spectra. In order to study this effect, we implement ionization and gas temperature evolution due to BHs in the version 4.5.1 of the public code CMBFAST (Seljak \& Zaldarriaga 1996; Seljak et al. 2003).

Fig.~\ref{fig:fig6} shows the temperature $-$ temperature (TT), polarization $-$ polarization (EE) and temperature $-$ polarization (TE) spectra of the CMB in the case in which the contribution from BHs is accounted for. In particular the `extreme' case (IMBH-6\%+MC01, solid line) and the 'fiducial' one (IMBH-3\%+PL1, dashed line) are shown. They are also compared with the spectra obtained without contributions from stars and/or BHs (thin dot-dashed line) and with the spectra derived assuming Thomson optical depth $\tau{}_e=0.09$ and a sudden reionization at $z\simeq11$, consistent with the 3-yr {\it Wilkinson Microwave Anisotropy Probe} ({\it  WMAP}) data.

As one can expect, no significant differences appear between the four cases in the TT spectrum. The 'fiducial' and the 'extreme' BH model differ from the thin dot-dashed line both in the EE and in the TE spectra, at low multipoles ($l\lesssim{}20$). However, the contribution of BHs to the TE and EE spectra, even in the extreme case (IMBH-6\%+MC01), is smaller than (or comparable to) the best fit of the 3-yr {\it WMAP} data (dotted line). Thus, all the scenarios considered in this paper (even IMBH-6\%+MC01) do not violate the limits posed by CMB observations. 

Furthermore, the Thomson optical depth which can be directly derived
from the ionization history shown in Fig.~\ref{fig:fig4} is
$\tau{}_e<0.07$ ($\tau{}_e=0.027$ and 0.064 in the 'fiducial' and
'extreme' case, respectively), smaller than the best fit to 3-yr {\it
WMAP} data ($\tau{}_e=0.09\pm{}0.03$, Spergel et al. 2007). Thus, our
BHs might give a partial contribution to the reionization, but are not
its exclusive source, in agreement with previous work (e.g. RO04;
Ricotti et al. 2005; Z07+).

\subsection{21 cm radiation}

\subsubsection{Basic definitions}

The spin temperature of the 21-cm transition can be written as (see
e.g. Field 1958, 1959; Kuhlen, Madau \& Montgomery 2006; Vald\`es et
al. 2007; Z07+)
\begin{equation}
T_{spin} = {{T_* + T_{CMB} + (y_k + y_{\alpha})
    T_k}\over{1+y_k+y_\alpha}},
\label{eq:tspin}
\end{equation}
where $T_*\equiv0.068\;{\rm K}$ corresponds to the 21-cm transition
energy, $T_{CMB}$ is the CMB temperature, $T_k$ is the IGM kinetic
temperature, and $y_k$ and $y_{\alpha}$ are the kinetic and Lyman
$\alpha$ coupling terms, respectively.

The kinetic coupling term is
\begin{equation}
y_k = {{T_*}\over{A_{10}\,T_k}}\,(C_H + C_e + C_p),
\label{eq:ykin}
\end{equation}
where $A_{10}\simeq2.85\times10^{-15}\;{\rm s^{-1}}$ is the Einstein
spontaneous emission rate coefficient (Wild 1952), and $C_H$, $C_e$ and
$C_p$ are the de-excitation rates due to neutral H, electrons and
protons, respectively. They are given by the fitting formulae from
Kuhlen et al. 2006 (see also Field 1958, 1959; Smith 1966; Allison \& Dalgarno
1969; Zygelman 2005):
\begin{eqnarray}
C_H & \simeq & 3.1\times10^{-11}
      \left({{T_k}\over{1\,{\rm K}}}\right)^{0.357}
      e^{-{{32\,{\rm K}}\over{T_k}}}\ {\rm s^{-1}}\\
\label{eq:kcoupling_H}
C_e & \simeq & n_e\, \gamma_e\\
\label{eq:kcoupling_e}
C_p & \simeq & 3.2\, (n_p/n_H)\, C_H,
\label{eq:kcoupling_p}
\end{eqnarray}
where $n_H$, $n_e$ and $n_p$ are the neutral H, electron and proton
number densities, and
\begin{equation}
{\rm log_{10}}{{\gamma_e}\over{1\;{\rm cm^3\,s^{-1}}}} \simeq
-9.607 + 0.5 {\rm log_{10}}{{T_k}\over{1\;{\rm K}}}
\exp{\left[{{1\over{1800}}
\left({{\rm log_{10}}{{T_k}\over{1\;{\rm K}}}}\right)^{4.5}}\right]}.
\label{eq:gammae}
\end{equation}

\begin{figure}
\center{{
\vspace{-0.5truecm}
\epsfig{figure=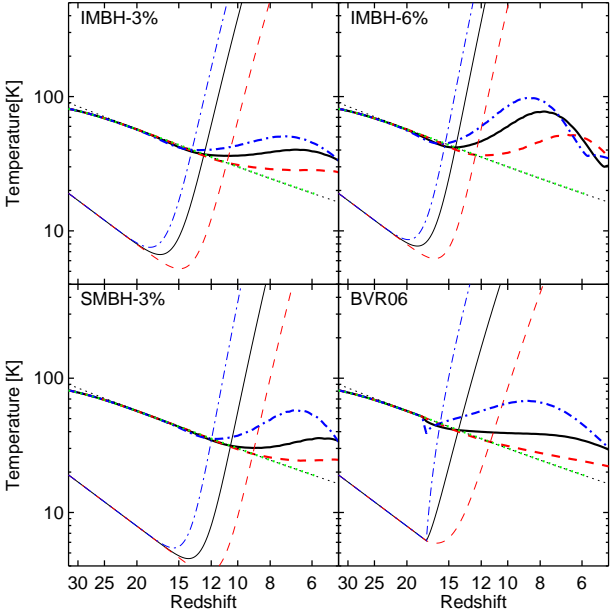,height=8.7cm}
}}
\caption{Redshift evolution of the neutral H spin temperature. The
order of the panels is the same as in Fig. \ref{fig:fig2}; the dotted
line represents the CMB temperature, thick lines represent the spin
temperature, and thin lines represent the IGM temperature. Continuous
lines refer to models with a PL1 SED, dashed lines to models with a
SOS1 SED, and dot-dashed line to models with a MC01 SED.}
\label{fig:fig7} 
\end{figure}

The Lyman $\alpha$ coupling term is given by
\begin{equation}
y_\alpha = {{16\pi}\over{27 A_{10}}}\, {T_*\over T_k}\,
{{\pi\,e^2}\over{m_e\,c}}\, f_{12}\, J_0,
\label{eq:yalpha}
\end{equation}
where $e$ and $m_e$ are the electron charge and mass,
$f_{12}\simeq0.416$ is the oscillator strength of the Lyman $\alpha$
transition, and $J_0$ is the intensity of Lyman $\alpha$ photons which
are due to collisional excitations from thermal electrons, to hydrogen
recombinations, and to collisional excitations from X-ray energy
absorption. $J_0$ is then
\begin{equation}
J_0 = {{hc}\over{4\pi\,H(z)}}
\left[{n_e n_H \gamma_{e,H} + n_e n_p \alpha_{2^2P}^{eff} + 
{{n_B \epsilon f_{exc}}\over{h\nu_\alpha}}}\right],
\label{eq:lyalphabkg}
\end{equation}
where $\gamma_{e,H}\simeq2.2\times10^{-8} e^{-(118400\,{\rm
K})/T_k}\;{\rm cm^3\,s^{-1}}$ is the collisional excitation rate of
neutral H atoms by electron impacts,
$\nu_\alpha\sim2.46\times10^{15}\;{\rm Hz}$ is the Lyman $\alpha$
frequency, and $\alpha_{2^2P}^{eff}$ is the
effective recombination coefficient to the $2^2P$ level (including
recombinations to the $2^2P$ level, plus recombinations to higher levels
that end up in the $2^2P$ level through all possible cascade paths). We
adopted a simple fit to the Pengelly (1964) results for
$\alpha_{2^2P}^{eff}$, assuming case A recombinations\footnote{Results
for case B recombinations differ only slightly.}:
\begin{equation}
\alpha_{2^2P}^{eff}(T_k)\simeq 1.67\times10^{-13}\,
T_4^{-0.91-(2/75){\rm log_2}{T_4}},
\label{eq:alpha22Peff}
\end{equation}
where $T_4 = T_k /(10^4\,{\rm K})$.

Once the spin temperature is known (from eq. \ref{eq:tspin}), it is
convenient to express the resulting 21-cm radiation intensity as the
differential brightness temperature between neutral hydrogen and the
CMB, which is an observable quantity:
\begin{equation}
\delta T_b \simeq {{T_{spin}-T_{CMB}}\over(1+z)}\,\tau_{21} (1+\delta_\rho),
\label{eq:deltatb}
\end{equation}
where $\delta_\rho \equiv (\rho-\bar\rho)/\bar\rho$ is the cosmological
density contrast in the considered region (here we will consider only
the case $\delta_\rho=0$), and $\tau_{21}$ is the IGM optical depth at
an observed wavelength of $21(1+z)\;{\rm cm}$,
\begin{equation}
\tau_{21} \simeq {{3 c^3 h A_{10}}\over
{32\pi\, k_B \nu_{21}^2 H(z)}}\,{n_H\over T_{spin}},
\label{eq:tau21}
\end{equation}
where $k_B$ is the Boltzmann constant, and $\nu_{21}\simeq
1.421\times10^9\;{\rm Hz}$ is the (rest-frame) frequency of the 21-cm line.


\begin{figure}
\center{{
\vspace{-0.5truecm}
\epsfig{figure=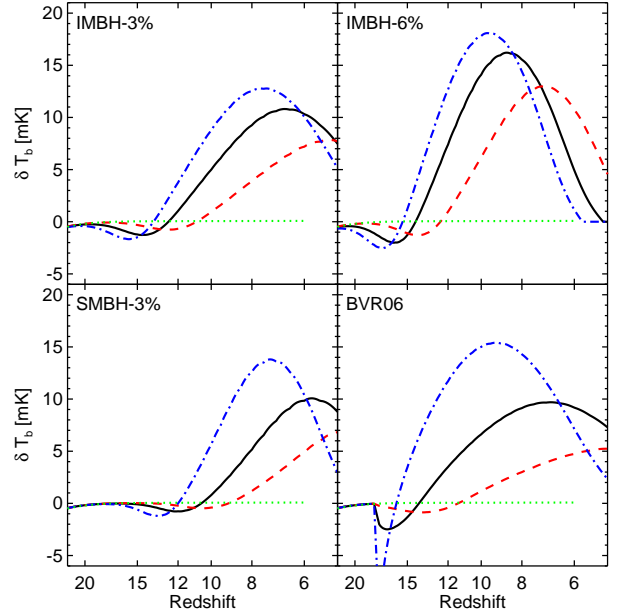,height=8.7cm}
}}
\caption{Redshift evolution of the brightness temperature difference
with respect to the CMB $\delta T_b$. The order of the panels and the
meaning of the various line types are the same as in
Fig. \ref{fig:fig2}, except for the dotted line, which represents
$\delta T_b$ for a model without any BH emission.}
\label{fig:fig9} 
\end{figure}

\subsubsection{Results for pure BH coupling}

Fig.~\ref{fig:fig7} shows the redshift evolution of $T_{spin}$,
under the assumption that only the radiation produced by BHs is
important.

In all these models, $T_{spin}$ remains very close to $T_{CMB}$ (and to
the predictions of models with no BH emission) until $z\sim{}9-15$,
i.e. until $T_k$ finally becomes much larger than $T_{CMB}$. After that, the
difference between $T_{spin}$ and $T_{CMB}$ becomes significant, and in
models with strong BH emission it can amount to $\sim 90\;{\rm K}$. Apart
from the amplitude of this maximum difference, the strength of BH emission
also influences the redshift when it is reached: in models
with weak BH emission (e.g. most models where a SOS1 SED is assumed),
$T_{spin}-T_{CMB}$ keeps increasing, and is largest at the lowest
considered redshift (though this maximum is quite low); whereas in
models with strong BH emission (e.g. all those where a MC01 SED is
assumed) $T_{spin}-T_{CMB}$ reaches a relatively high maximum at
$z\sim{}6-9$, slowly decreasing afterwards. The main reason is that in
models with high BH emission the ionized fraction easily reaches the
regime (at $x_H\gtrsim 0.1$) where $f_{exc}$ (and $J_0$ and $y_\alpha$
with it, as $J_0$ is dominated by the term due to collisional
excitations from X-ray absorption) starts dropping very fast, rather
than being approximately constant (see fig. 4 of Shull \& Van Steenberg
1985).

In Fig.~\ref{fig:fig9} we show the corresponding evolution of the
differential brightness temperature $\delta T_b$. Such evolution
essentially mirrors the one of $T_{spin}-T_{CMB}$: it remains close to 0
until $z\sim{}9-15$, and then starts growing, reaching maxima between
$\sim{}5$ and $\sim{}18\;{\rm mK}$, depending on the strength of the BH
emission. Again, in models with weak BH emission the maximum is reached
at the lowest considered redshift, whereas in the other models it is
reached at $z\sim{}6-9$. The main difference with the evolution of
$T_{spin}-T_{CMB}$ is that the decline after the maximum is faster,
since the high IGM ionization level in models with strong BH emission
reduces also $\tau_{21}$.

It must be remarked that such an evolution of $\delta T_b$ in the
neutral patches of the Universe at $z\lesssim 12$ should be detectable
with the new generation of radio experiments, such as LOFAR, MWA, 21CMA
and SKA\footnote{http://www.lofar.org\\
http://www.haystack.mit.edu/ast/arrays/mwa/ \\
http://21cma.bao.ac.cn/\\http://www.skatelescope.org}. For example,
LOFAR will probe the 21~cm emitted from the IGM in the redshift range of
6--11.5 and will be sensitive to scales from a few arcminutes up to few
degrees and will be able to statistically detect the 21~cm brightness
temperature down to $\approx$5 mK (de Bruyn, Zaroubi \& Koopmans 2007,
de Bruyn et al. 2008).
 However, we stress that these effects are
 observable only before the end of the reionization epoch (see Section 5.1).

\subsubsection{Results for BH and stellar coupling}

In the previous Subsection we considered the evolution of the 21-cm
emission under the effects of BH emission only. But it is largely
believed that stellar emission played a fundamental role in the
evolution of the primordial Universe: for example, most models of
reionization (e.g. RO04, and references therein) assume that the stellar
contribution was dominant over the one from BHs. This is supported by
observations of the unresolved X-ray background, whose level is
difficult to reconcile with the hypothesis that reionization is due to
BH emission (e.g. Dijkstra et al. 2004). Furthermore, our models
\emph{do} require the presence of stellar radiation, as even the
`extreme' one (IMBH-6\%+MC01) is unable to reionize the Universe before
$z\sim 6$, and is therefore incompatible with observations of quasars
and Lyman $\alpha$ emitters at $z\sim 6-7$ (Becker et al. 2001;
Djorgovski et al. 2001; Fan et al. 2001, 2002, 2003; White et al. 2003;
Kashikawa et al 2006; Iye et al. 2006; Ota et al. 2007).

Since we are looking at the evolution of the IGM in regions which are
quite removed from BHs (and, consequently, from the bulk of stellar
emission) and are reionized late\footnote{It is natural to wonder down to which redshift such neutral regions actually exist. Here, we will simply assume that they survive down to $z\sim5$, and look into their properties. Such hypothesis will be discussed in Section 5.1.}, the omission of the stellar
contribution from our calculations is mostly justified.
In fact, the neutral IGM we are considering is almost
perfectly transparent to radiation with frequencies below the H
ionization threshold ($13.6\;{\rm eV}$): such photons can travel
cosmological distances, but are unable to significantly affect the
IGM. On the other hand, the ionizing photons emitted from stars are
typically absorbed at the edge of the ionized regions around stellar
sources, since they are not energetic enough to cross significant
distances in a neutral IGM: for this reason, their effects are purely
local.

\begin{figure}
\center{{
\vspace{-0.5truecm}
\epsfig{figure=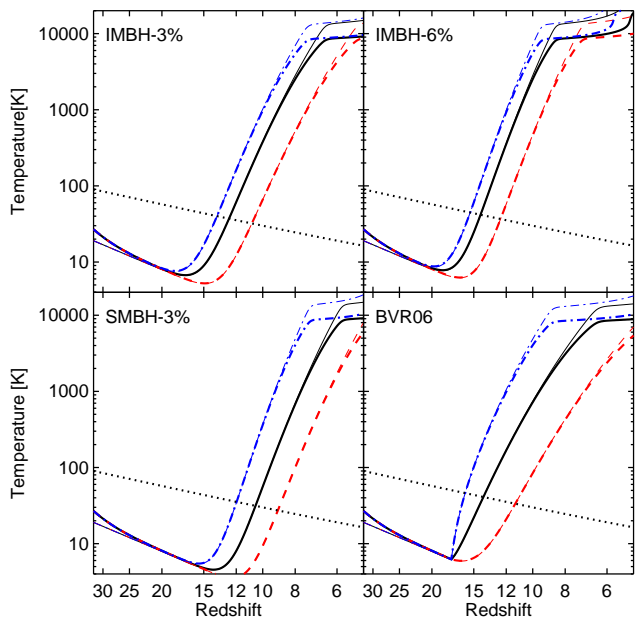,height=8.7cm}
}}
\caption{Redshift evolution of the neutral H spin temperature, when the
Ly$\alpha$ coupling due to stellar radiation (but not the stellar
radiation heating effects) is kept into account. The order of the panels
is the same as in Fig. \ref{fig:fig2}. The dotted line represents the
CMB temperature, thick lines represent the spin temperature, and thin
lines represent the IGM temperature. Continuous lines refer to
models with a PL1 SED, dashed lines to models with a SOS1 SED, and
dot-dashed line to models with a MC01 SED.}
\label{fig:fig10} 
\end{figure}


\begin{figure}
\center{{
\vspace{-0.5truecm}
\epsfig{figure=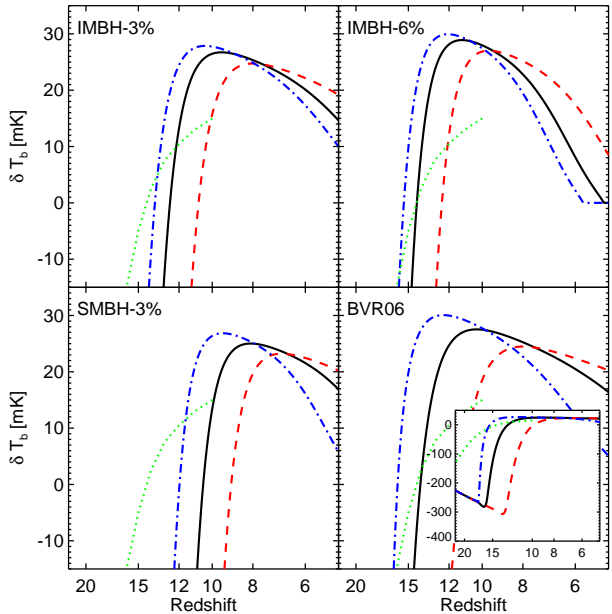,height=8.7cm}
}}
\caption{Redshift evolution of the brightness temperature difference
with respect to the CMB $\delta T_b$, when the Ly$\alpha$ coupling due
to stellar radiation is kept into account. The order of the panels and
the meaning of the various line types are the same as in
Fig. \ref{fig:fig2}. The insert in the bottom-right panel shows the
same quantities (for the BVR06 case; the other cases are qualitatively
similar) on a much wider $\delta T_b$ scale. The dotted line stopping at
$z=10$ comes from Fig. 5 (bottom panel, solid line) of CS07, and shows
$\delta T_b$ for a model with stellar Lyman $\alpha$ coupling and
heating, but no BH emission. This line represents an upper limit on
$\delta T_b$ in the absence of BH heating.}
\label{fig:fig12} 
\end{figure}

Lyman $\alpha$ photons are the only relevant exception. In fact,
although the Lyman $\alpha$ cross section is very high, such photons can
scatter many times before exiting the resonance; more importantly, the
redshifting of photons with energies slightly higher than $10.2\;{\rm
eV}$ `into' the resonance ensures a roughly uniform Lyman $\alpha$
radiation field also in neutral regions.

Ciardi \& Salvaterra (2007; hereafter CS07) found that the Lyman
$\alpha$ radiation field can moderately heat the IGM: the heating rate
taken from their models dominates over that of all our models at
$z\gtrsim15$, and takes the IGM temperature to $\gtrsim 30\;{\rm K}$ at
$z\sim15-20$. On the other hand, at $z\lesssim 10$ the Lyman $\alpha$
heating rate should be much smaller than those of our
models\footnote{The plots in CS07 actually stop at $z=10$; but it is
pretty clear that in their model the IGM temperature is growing at a
much slower rate than in our models. It is also worth noting that
some of the CS07 assumptions (e.g. the values of the parameters
$f_{gas}$ and $f_*$) are quite extreme, and would result in a very early
complete reionization. More realistic assumptions would
result in a significant delay in the rise of $T_{spin}$.}. More
importantly, CS07 find that for $z\lesssim 27$ the intensity
$J_{\alpha,*}$ of the Lyman $\alpha$ background is much higher than the
level [$J_{\alpha,coupling} \sim 10^{-22} (1+z)\;{\rm erg\, cm^{-2}\,
s^{-1}\, Hz^{-1}\, sr^{-1}}$] which should couple $T_{spin}$ to $T_k$
rather than to $T_{CMB}$ (see Ciardi \& Madau 2003).

In our case, it is reasonable to neglect the heating effects of the
Lyman $\alpha$ background, although this will lead us to somewhat
underestimate the IGM temperature at $z\gtrsim{}15$. But it is very
important to add the effects of the Lyman $\alpha$ background to our
estimation of $T_{spin}$ (and $\delta T_b$).

This can be done very easily by modifying eq.~(\ref{eq:tspin}) into
\begin{equation}
T_{spin} = {{T_* + T_{CMB} + (y_k + y_{\alpha} + y_{\alpha,*})
    T_k}\over{1+y_k+y_\alpha+y_{\alpha,*}}},
\label{eq:tspinstar}
\end{equation}
where $y_{\alpha,*}$ accounts for the additional coupling due to the
Lyman $\alpha$ background due to the stars\footnote{Also BHs produce a
Lyman $\alpha$ background; but its intensity is much lower than the one
due to stars, and the corresponding coupling term is always much smaller
than $y_{\alpha}$.}, and is approximately given by (see CS07):
\begin{equation}
y_{\alpha,*}\sim 10^9 {{J_{\alpha,*}\,T_*}\over{A_{10}\,T_k}}.
\label{eq:yalphastar}
\end{equation}

After approximating $J_{\alpha,*}$ with the expression
\begin{equation}
{{J_{\alpha,*}(z)}\over{{\rm
      erg/(cm^2\,s\,Hz\,sr)}}} =
\left\{{
  \begin{array}{ll}
    0                    & z \ge 30\\
    10^{-18-[0.1(z-10)]} & 30 > z > 10\\
    10^{-18}             & z \le 10,\\
\end{array}}\right .
\label{eq:jalphastar_cs}
\end{equation}
which is a moderate underestimate of the $J_{\alpha,*}$ curves shown in
fig.~1 of CS07, we have recalculated the
evolution of $T_{spin}$, and $\delta T_b$. The
results are shown in
Figs.~\ref{fig:fig10}~and~\ref{fig:fig12}.

In this case, $T_{spin}$ (Fig.~\ref{fig:fig10}) is almost perfectly
coupled to the kinetic temperature, and the difference
$T_{spin}-T_{CMB}\simeq T_{spin}$ easily reaches the $10^3-10^4\;{\rm
K}$ range. Also $\delta T_b$ (Fig.~\ref{fig:fig12}) is affected. Here we
focus on the relatively low redshifts which will be explored by 21-cm
experiments (e.g. LOFAR), where the effects of BH emission lead to
differential brightness temperatures which can reach $20-30\;{\rm mK}$
at redshifts $\sim 8-15$. Instead, at high redshifts (say,
$z\gtrsim{}15$) $\delta T_b$ can reach very high negative values (in the
$-200$ to $-300\;{\rm mK}$ range); but in such redshift range the
results of CS07, predicting a minimum value of $\delta T_b\sim-170\;{\rm
mK}$ at $z\sim24$ are likely more correct because they include also the
the heating effects of the stellar Lyman $\alpha$ background.

We point out that our results, especially those about $\delta T_b$,
depend only weakly from the very high level of $J_{\alpha,*}$ given
in the CS07 paper: the effects of lowering $J_{\alpha,*}$ to a more
realistic level, e.g. a fraction $0.1$ (or even $0.01$) of the
amount given by eq. (\ref{eq:jalphastar_cs}) are a certain reduction
(from $\sim8-10000\;{\rm K}$ to $3-5000\;{\rm K}$) of the level where
$T_{spin}$ reaches a low-redshift `plateau', and a much smaller change
in the evolution of $\delta T_b$. Then, our predictions about $\delta
T_b$ observations are quite independent from the assumptions of
CS07. Instead, for the model where no BH feedback is included, a
reduction by a similar factor in the Lyman-$\alpha$ heating rate in the
CS07 models would result in a much lower $\delta T_b$ value than shown
by the dotted curve in Fig.~\ref{fig:fig12}.

\section{Influence on structure formation}

In the previous section we have shown that the energy input from BHs can
substantially heat the IGM. In turn, this is likely to affect the
formation of galaxies: as the cosmological Jeans mass depends on
$T_k^{3/2}$, the baryonic component of small fluctuations might become
unable to collapse and form stars because of the temperature
increase. But the effects of BH radiation are not limited to the
heating, since the increase in the H ionized fraction also enhances the
formation of H$_2$, which is the most important coolant in metal-free
gas at temperatures $\lesssim 10^4\;{\rm K}$: such enhancement would
facilitate the formation of stars within small halos.  Then, we
investigated the influence of BH energy input on structure formation
with a method which accounts for such competing effects, and which was
already employed in the RMF07 and R07 papers.

We used the full code (instead of the simplified version used in the
previous Section) described in Section 3, in order to follow the
evolution of spherically symmetric halos of different masses,
virializing at different redshifts. Such evolution took into account all
the physics included in the simplified version we already described,
plus the treatment of gravity, of the hydrodynamical evolution of the
gas, and of the dissociation of H$_2$ molecules due to Lyman-Werner
($11.2\,{\rm eV}\le h\nu\le 13.5\,{\rm eV}$) photons emitted by
BHs\footnote{For this last effect we use the reaction rate given by Abel
et al. 1997 (reaction 27); the flux of photons at $12.87\,{\rm eV}$ was
obtained through the formalism described in Section 2.1, but assuming
that photons at frequencies corresponding to the lines of the Lyman
series of hydrogen were completely absorbed. No stellar emission was
assumed.}. Dark matter (DM) gravitational effects are included as
described in Sec. 2.1.3 of R07: the DM final density profile is assumed
to be a truncated isothermal sphere with $\xi=0.1$ (i.e., the final core
radius is assumed to be 1/10 of the virial radius).

\subsection{Critical mass}

\begin{figure}
\center{{
\vspace{-0.5truecm}
\epsfig{figure=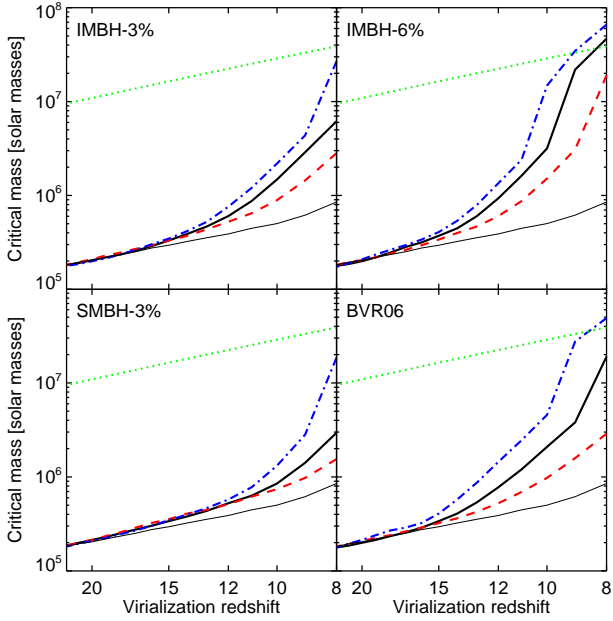,height=8.7cm}
}}
\caption{Redshift evolution of the critical mass. The order of the
panels and the meaning of the various line types are the same as in
Fig. \ref{fig:fig2}, except for the thin continuous line, which represents
$M_{crit}$ for a model without any BH emission, and for the dotted line,
representing the mass $M_H$ of halos with virial temperature
$T_{vir}=10^4\;{\rm K}$.}
\label{fig:fig13} 
\end{figure}

As in the RMF07 and R07, we classified halos as
collapsing if they reach a maximum density larger than
$\rho_{coll}=1.67\times10^{-19}\,{\rm g\,cm^{-3}}\simeq10^5 m_H\,{\rm
cm^{-3}}$ (a value high enough to suggest that the formation of a
luminous object is well under way) in less than an Hubble time after
their virialization (at $z_{vir}$), i.e. at a redshift
$z_{coll}\gtrsim[0.63(1+z_{vir})]-1$.

This classification criterion is roughly comparable to the collapse
criterion of Tegmark et al. (1997): in analogy with such paper (and
with RMF07 and R07), we define the critical mass
$M_{crit}(z_{vir})$ as the minimum mass of a collapsing halo virializing
at $z_{vir}$.

In Fig. \ref{fig:fig13} we compare the evolution of $M_{crit}$ which is
obtained for each of our BH models with the
same evolution in the unperturbed ($\epsilon=0$ at all redshifts)
case, and with the evolution of the mass
\begin{equation}
M_H(z_{vir})\simeq 1.05\times10^9\,\Msun\,(1+z_{vir})^{-3/2}
\label{eq:mtvir10000}
\end{equation}
of halos with a virial temperature $T_{vir}=10^4\;{\rm K}$ (assuming a
mean molecular weight $\mu{}= 1.23$, as appropriate for a neutral medium),
above which the cooling due to atomic H becomes dominant.

The BH energy injection has negligible effects upon $M_{crit}$ for
$z_{vir}\gtrsim 15$, but its effects become increasingly important at
later times: at $z=10$ the BH energy input increases $M_{crit}$ by a
factor between $1.8$ (SMBH-3\%+SOS1 model) and $40$ (IMBH-6\%+MC01
model). At lower redshifts the BH effects are even larger: in the models
with the strongest BH feedback, $M_{crit}$ can become $\gtrsim M_H$,
although the onset of atomic cooling slows down the increase of
$M_{crit}$: in such models, BH feedback prevents the formation of stars
inside mini-halos cooled by molecules at $z\lesssim{}9$.

\begin{figure}
\center{{
\vspace{-0.5truecm}
\epsfig{figure=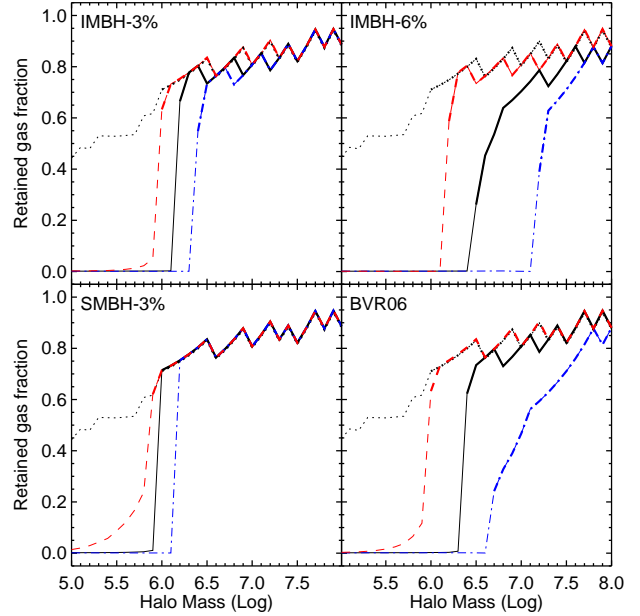,height=8.7cm}
}}
\caption{Fraction of gas retained by a halo at the end of our
simulations in halos virializing at $z_{vir}=10$, as a function of the
total halo mass. The order of the panels and the meaning of the
continuous, dashed and dot-dashed lines are the same as in
Fig. \ref{fig:fig2}; the dotted lines refer to the unperturbed
model. The thickness of the line indicates whether a certain halo mass
is below (thin line) or above (thick line) the critical mass. The seesaw
behavior for high masses is purely numerical (i.e. due to the discrete
number of shells).}
\label{fig:fig14} 
\end{figure}

\subsection{Gas retention}

\begin{figure}
\center{{
\vspace{-0.5truecm}
\epsfig{figure=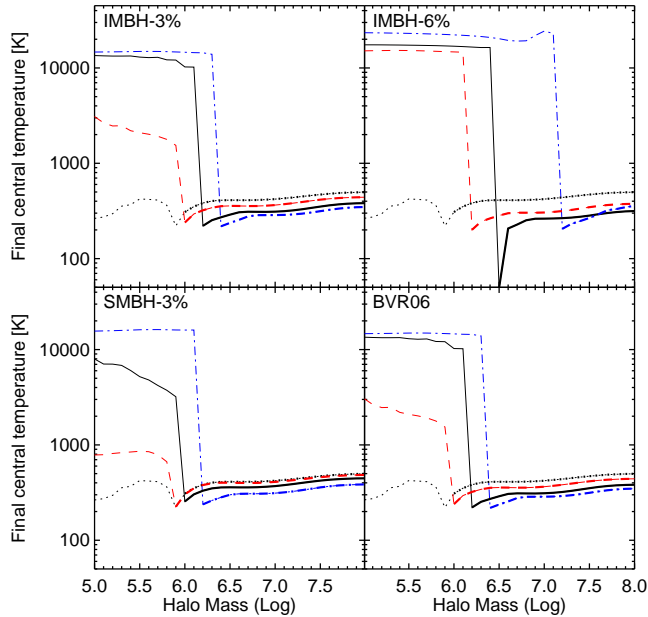,height=8.7cm}
}}
\caption{Central gas temperature at the end of our simulations in halos
virializing at $z_{vir}=10$, as a function of the total halo mass. The
order of the panels and the meaning of the continuous, dashed and
dot-dashed lines are the same as in Fig. \ref{fig:fig2}; the dotted
lines refer to the unperturbed model. The thickness of the line
indicates whether a certain halo mass is below (thin line) or above
(thick line) the critical mass.}
\label{fig:fig15} 
\end{figure}

RMF07 suggested that one possible feedback effect of the energy input
from decaying/annihilating DM particles is to reduce the amount of gas
which actually ends up within the potential wells of virialized
halos. As the feedback effects of BHs are much stronger than those of
DM decays and annihilations, we looked at whether these same effects are
important in our simulations.

To this purpose, we define $f_{ret}$ as the ratio of the mass of gas which is
retained inside the virial radius of a halo  (at the time when our simulations
are stopped) with respect to the baryonic mass expected from cosmology. For a
halo with total mass $M_{halo}$ and baryonic mass $M_{gas}$:
\begin{equation}
f_{ret} = {M_{gas}\over M_{halo}}\,{}{\Omega_m\over\Omega_b}.
\end{equation}

In Fig.~\ref{fig:fig14} we show the dependence of $f_{ret}$ upon
$M_{halo}$, for halos virializing at $z_{vir}=10$ and for all our BH
models, plus the unperturbed case.


Generally, models with BH feedback exhibit a sharp transition at
$M_{halo}\simeq M_{crit}$, going from $f_{ret}\sim0$ to
$f_{ret}\gtrsim0.7$, whereas in the unperturbed case $f_{ret}$ increases
quite smoothly from $\sim 0.4$ to $\sim 0.9$ (this last value is in good
agreement with numerical simulations by Crain et al. 2007). This
threshold effect is due to the hydrodynamical effects of the BH heating,
combined with the depth of the DM potential well. In fact, the heating
induced by the BHs amplifies the pressure gradients and tends to prevent
the gas from falling inside the DM potential well. If the DM potential
well is below a certain critical value (i.e. if $M_{halo}\lesssim
M_{crit}$), the final gas over-density is generally $\lesssim{}10$, whereas
the DM over-density is $\gtrsim{}1000$.  On the other hand, if the gravity
of a halo is strong enough (i.e. if $M_{halo}\gtrsim M_{crit}$), the
heating induced by the BHs cannot counteract the gravitational pull, and
the halo will retain most of its gas, which will cool, collapse and form
luminous objects.

This is confirmed by Fig.~\ref{fig:fig15}, where we show the final
temperature of the gas at the centre of halos virializing at
$z_{vir}=10$, as a function of the halo mass. In halos with mass $\le
M_{crit}$ the gas temperature is much higher, if BH heating is present,
than in the unperturbed case, and it is close to the temperature $T_k$
of the IGM ($1000-10000\;{\rm K}$). For masses $\ge M_{crit}$ the final
temperature in presence of BH heating is similar to the unperturbed case
($200-400\;{\rm K}$, much lower than the temperature of the surrounding
IGM), as the gas in the centre of the halo was able to condense and
cool. The transition in the case of the gas temperature is even sharper
than in the case of $f_{ret}$, probably because the density dependence
of the cooling rate\footnote{In most of the regimes we are considering,
the cooling rate is due to H$_2$ molecules. An increase in density
results in both an increase in the cooling rate per molecule (which is
$\propto\rho$) and an increase in the abundance of molecules.} will lead
to a `runaway' cooling as soon as the density starts to increase.

Fig.~\ref{fig:fig14} also shows that in models with strong BH feedback
(IMBH6\%+PL1, IMBH6\%+MC01, and BVR06+MC01) the transition from
$f_{ret}\sim0$ to $f_{ret}\sim0.8$ is not as sharp as in the other cases
we consider (remarkably, such difference is not present in the
temperature plots of Fig.~\ref{fig:fig15}). In such models, halos with
masses in the range $M_{crit}\le M_{halo} \lesssim 5 M_{crit}$ are
relatively poor in gas, despite being able to form luminous objects at
their centre. Such a luminous but gas-poor halo population starts
developing at $z\sim 12$, and becomes increasingly important when lower
redshifts are considered: for instance, at $z\sim{}8$ this population is
present also in models with intermediate BH feedback, and can span a
factor of $\sim 10$ in mass. If such objects actually exist and survive
until present, they should be characterized by a high $M/L$ ratio, a 
low gas content, and a mass $\sim 10^8\Msun$. Such properties remind us
of the dwarf spheroidal galaxies of the Local Group (see Mateo
1998), although it might just be a coincidence. Further investigation is
needed to address this issue.

\section{Discussion}

\subsection{Do neutral regions exist below redshift 11?}


The 3-yr WMAP results (Spergel et~al. 2007) for the electron
scattering optical depth can be interpreted as indicating a sudden
reionization at $z\simeq11$. In such a scenario, neutral regions
essentially cease to exist as soon as reionization happens, and the
effects of BHs at $z\lesssim 11$ would become negligible:
\begin{itemize}
\item{}{The 21 cm brightness temperature differences would be severely
quenched because of the lack of neutral H ($\delta T_B$ is
proportional to the density of neutral H atoms). Even if it were
not, it is reasonable to expect that the reionization takes the IGM
temperature $T_k$  to $\sim 10^4\,{\rm K}$, and the BH heating would
be unable to drastically change $T_k$; the changes in $T_{spin}$
would be even smaller.}
\item{}{The high IGM temperature we just mentioned would probably have
important effects on structure formation; but that is a feedback
effect from {\it stellar} sources, rather than from BHs.}
\end{itemize}
In short, the effects of BHs can be clearly observed only at redshifts
before the end of reionization process. In the case of a sudden
reionization at $z\simeq11$ they would become extremely difficult to
detect, except perhaps in our models with the strongest BH feedback.

However, the sudden reionization scenario appears unrealistic. In fact,
practically all the theoretical models predict that the reionization
process is quite extended in time. In particular, the most recent
numerical simulations (e.g. Iliev et al. 2007; Santos et al. 2007; Zahn
et al. 2007; Mesinger \& Furlanetto 2007) essentially agree in the
prediction that the end of the overlap phase (i.e., the time when the
volume filling factor of neutral regions becomes negligible) is at
redshift $6.5\lesssim z_{overlap} \lesssim 8$. 
Fig. 3 of Santos et al. 2007 is particularly useful
for our purposes, since it includes not only the evolution of the
volume-averaged ionization fraction (solid line), but also a similar
curve where complete ionization is assumed within the ionized
`bubbles' (dashed line): it is quite reasonable to expect the volume
filling-factor of neutral regions to drop below $\sim 0.1$ when such
curve exceeds $0.8-0.9$, i.e at $z\lesssim 7-8$.

Such behavior is broadly consistent also with analytical models such as
the one presented in Choudhury \& Ferrara 2006 (in their fig.~1a the
volume-averaged neutral fraction  goes below $0.1$ already at $z\simeq9$,
but declines below $0.01$ only at $z\lesssim6$). It is also important to
point out that such low values for $z_{overlap}$ are usually obtained in
models based on the 3-yr WMAP data, whereas models based upon 1-yr WMAP
data (Kogut et al. 2003; Spergel et al. 2003) lead to significantly
higher values of $z_{overlap}$ (for instance, see Iliev et al. 2007,
which presents the results of simulations based on both sets of
parameters).

Furthermore, observations of the Gunn-Peterson troughs in the spectra of
quasars at $z\gtrsim6$ (see e.g. Fan et. al. 2002; Fan et al. 2006), and
measurements of the evolution of the density of bright Lyman $\alpha$
emitters at $5.7\lesssim z \lesssim 7$ (Kashikawa et al. 2006; Iye et
al. 2006; Ota et al. 2007) might hint that we are actually observing the
final stages of overlap; but the interpretation of the data is difficult
and the issue is still under intense debate (see e.g. Malhotra \& Rhoads
2004; Fan, Carilli \& Keating 2006; Mesinger \& Haiman 2007; Dijkstra,
Wyithe \& Haiman 2007).

In short, it seems reasonably likely that the volume filling factor of
neutral regions remained significant ($\gtrsim 0.1$) at least until
redshift $7-8$, and maybe even at lower redshifts; nonetheless, it
is also possible (e.g. if the 3-yr WMAP results are underestimating
$\tau_e$) that a dearth of neutral regions at $z\lesssim 10$ will
prevent the detection of the effects we discuss.

\subsection{Comparison with previous works about BH feedback}
Our analysis of feedback effects from high-redshift BHs has many links
with the former study by RO04 (and also with Ricotti et
al. 2005). However, there are some crucial differences between the
assumptions in the two sets of models, which lead to important
differences in the results.
\begin{itemize}
\item{}{The most important difference is likely to be in the growth
histories of the BH densities. All of the RO04 models reach
$\rho_{BH}\gtrsim 10^5\;{\rm \Msun\,Mpc^{-3}}$ (in their notation,
$\omega_{BH}\sim 1.7\times 10^{-5}$; cfr. the lower panel of their
fig. 2) at redshifts $\ge 15$, whereas at $z=15$ none of our
models exceeds $\rho_{BH}=10^3\;{\rm \Msun\,Mpc^{-3}}$. This difference
becomes less important when going to lower redshifts. The IMBH-6\%
growth history actually overtakes the RO04 predictions at $z\lesssim
6-7$. But all the other growth histories we consider are
at most comparable to the RO04 models even at $z=5$.}
\item{}{We assume a constant duty cycle ($y=0.03,\ 0.06,\ {\rm or}\
0.10$, depending on the model), whereas in the RO04 models this
quantity strongly depends on redshift (see the bottom panel of their
fig.~3): it is assumed to be $1$ at high redshifts ($z\ge14$, $z\ge
19$, or $z\ge 24$), but rapidly falls to $10^{-3}$  when lower
redshifts are considered ($z\le13$, or $z\le8$).}
\item{}{RO04 restrict their analysis to an intrinsically absorbed
Sazonov et al. (2004) spectrum; their treatment of radiation transfer is
more detailed than in the present paper, but as they are not limiting
themselves to the neutral-IGM, their background spectrum is likely to
extend to lower energies than ours, resembling the thin solid line in
Fig.~\ref{fig:fig1}.}
\item{}{The RO04 models include also a stellar contribution.}
\end{itemize}

Because of all these differences, the RO04 models predict a much larger
energy injection into the IGM (at $z\gtrsim15$ the difference can easily
amount to a factor of $\gtrsim10^3$). At lower redshifts ($z\lesssim 8-9$)
such difference is erased (or even reverted), mostly because of the
reduction of the RO04 duty cycle.

Taking into account these differences, the results of the current paper
are reasonably consistent with those of RO04. In fig.~5 of RO04, the
ionized fraction and the IGM temperature are shown for different models.
Complete ionization is achieved already at $z\sim{}7-8$, while in our
models $x_{\rm HI}$ is always less than 1 at $z>6$ (but in most of our
models complete reionization is never reached). This difference is
simply explained by the presence of a stellar component in the RO04
semi-analytical model. For the same reason, IGM temperatures of 10$^4$ K
are reached at $z\sim{}8-10$ in our paper and at $z\sim{}20-25$ in
RO04. The Thomson optical depth derived by RO04 is
$0.1\lesssim{}\tau{}_e\lesssim{}0.2$, but a fraction
$\tau{}_e\sim{}0.06$ is due to stars. Thus, the contribution of BHs to
the Thomson optical depth in RO04 models is
$\tau{}_e\approx{}0.04-0.14$, which is consistent with our findings
($\tau{}_e\lesssim{}0.07$). Furthermore, RO04 aim to reproduce the
Thomson optical depth derived from 1-yr WMAP results
($\tau{}_e\simeq{}0.17\pm0.05$), which is considerably higher than in
the 3-yr WMAP measurements ($\tau{}_e\simeq{}0.09\pm0.03$).

Ricotti et al. (2005) also study the effects on the 21-cm line; but, in
their fig. 10, $\delta{}T_{\rm b}$ starts increasing already at
$z\sim{}20-25$, because of the strong
increase in the IGM temperature due to the BH emission. The predicted
peak in $\delta T_B$ is of the order of only a few mK, a factor of $\sim
10$ smaller than in our models. The low ($\lesssim 0.2$) neutral
fraction in their models is the likely cause of this discrepancy, as it
implies a low $\tau_{21}$ in eq. (\ref{eq:deltatb}).

\subsection{Other X-ray feedback mechanisms}

\begin{figure}
\center{{
\vspace{-0.5truecm}
\epsfig{figure=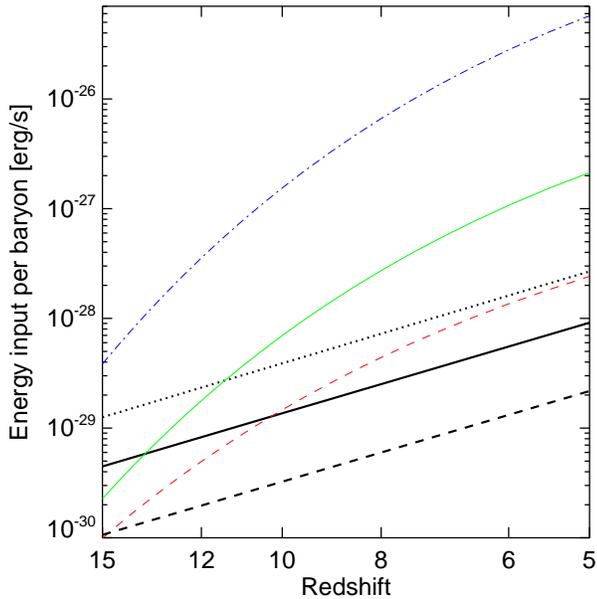,height=8.7cm}
}}
\caption{Redshift evolution of the total energy input per baryon due to
the background produced by X-ray emission associated with star formation
(thick lines; the two lines refer to power-law SEDs with different
photon index: $\Gamma=1.5$ for the dashed line, $\Gamma=2.0$ for the
solid line, and $\Gamma=2.5$ for the dotted line), assuming $f_X=1$. For
comparison, we show also the energy input in models where BH emission
was considered (IMBH-6\% + MC01: thin dot-dashed line; IMBH-3\%+PL1:
thin solid line; BVR06 + SOS1: thin dashed line).}
\label{fig:fig16} 
\end{figure}

Observations of local star-burst galaxies (Grimm, Gilfanov \& Sunyaev
2003; Ranalli, Comastri \& Setti 2003; Gilfanov, Grimm \& Sunyaev 2004)
find a correlation between star formation and X-ray
luminosity. As was noted in
Glover \& Brand~(2003), Furlanetto~(2006),
Pritchard~\&~Furlanetto~(2007) and Santos~et~al.~(2007),
it is reasonable to expect that also high redshift star formation is
associated with X-ray emission, although an unknown (and possibly
important) correction factor $f_X$ should be introduced to quantify the
differences between the local and the primordial environment.

The effects of such emission (and whether they can be distinguished from
the ones presented in this paper) will be thoroughly investigated in a
companion paper (Ripamonti et al., 2008 - in preparation). Here, we just
compare the energy input due to BHs with the one due to X-ray emission
associated with star formation. This was done by assuming that the SED
of such emission is a power-law with photon index in the
$1.5\le\Gamma\le2.5$ range, and that the star formation rate is similar
to the one shown in fig.~1b of Choudhury \& Ferrara (2006). In
Fig.~\ref{fig:fig16} we show that  in the lower redshift range we
consider ($z\lesssim10$) the expected energy input is only a fraction
of the BH contribution of most of our models, and is comparable only to
the BH model with the weakest feedback; at higher redshifts the
SF-associated X-ray emission might be more important or even dominant,
but the overall energy input is small.

In short, if the unknown factor $f_X$ is not much larger than 1, the
effects of the X-ray emission associated with star formation should be at
most comparable to the ones of the weakest of our BH models (such as the
BVR06+SOS1 case).

\section{Summary and conclusions}
We have examined how a population of accreting BHs might affect the
pre-reionization Universe, looking in particular at the effects upon the
neutral regions outside the first ionized `bubbles', where stellar
feedback is likely small. We explored a number of scenarios for the
growth of the cosmological BH density, and considered several possible
SEDs. Both of these components are in broad agreement with observational
constraints (e.g. the results about the X-ray background by Dijkstra et
al. 2004).

Our analysis started from how the energy input from the diffuse
radiation due to the BH population might affect the temperature and
ionization level of the IGM far-away from ionized regions where local effects 
are important. Given the Dijkstra~et~al.~(2004)
constraints, it is not surprising that BH emission in our models leads
only to partial ionization: the main effect of BH emission is then the
increase in the temperature of the IGM, which easily reaches levels
$\gtrsim 10^3\;{\rm K}$ in all the cases we have considered.

Then, we explored a number of possible indirect consequences of the
energy input:
\begin{itemize}
\item{}{CMB measurements appear unable to constrain any of our
models, since all of them comfortably fit observational constraints from
WMAP;}
\item{}{21-cm observations appear extremely promising, since in most of
the BH models the predicted $\delta T_b$ should be easily detectable
with the next generation of 21-cm experiments (e.g. LOFAR), especially
if stellar Lyman $\alpha$ coupling is really present;}
\item{}{the critical mass for halos to be able to cool, collapse and
form stars is significantly enhanced  at $z\lesssim 10$, and in
some of our models it becomes $\sim 100$ times larger than in the
unperturbed case. This allows star formation only in halos with virial
temperatures $\gtrsim 10^4\;{\rm K}$, i.e. prevents (or, in models with
weak feedback, significantly reduces) the formation of PopIII objects
for $z\lesssim 9$;}
\item{}{gas depletion might occur in the models with
intermediate-to-strong BH feedback, and for relatively low
virialization redshifts: halos with masses between $M_{crit}$ and
$3-10\,M_{crit}$ appear to be able to form stars at their centre, but
their baryonic fraction is considerably lower than the cosmological
average.}
\end{itemize}


The most relevant of our results appears to be the one about 21-cm
observations, since it might be falsified (or confirmed) by forthcoming
observations. To our knowledge, the only mechanism which should be able
to heat the IGM outside ionized regions in a comparable way is the X-ray
emission associated with star formation, as was proposed by Glover \&
Brand (2003), Furlanetto (2006) and Pritchard \& Furlanetto (2007). We
leave the detailed comparison between the two models to a future paper
(Ripamonti et al. 2008, in preparation), where we will also investigate
whether it is possible to distinguish between the two scenarios, e.g. by
using the spatial power-spectrum.

We stress that most of our conclusions strongly depend on the details of
the reionization process, and in particular on the survival of neutral
regions down to redshift $\sim{}7-8$. Recent simulations (e.g. Santos et
al.  2007) and observations (e.g. Fan et~al. 2002, Kashikawa
et~al. 2006) hint that the overlapping phase lasted for a long time and
suggest the existence of patches with a significant neutral fraction
even at $z\lesssim7$. However, the scenario of an earlier reionization
cannot be rejected at present: in such a case, BH signatures (such as
the effects on the properties of 21-cm radiation) become difficult or
impossible to detect. On the other hand, if the predictions of
simulations are correct, the effects of BH emission might enhance the
21-cm contrast between neutral and ionized patches, improving our
capability of studying the $z\sim{}7-12$ Universe, and providing
important information on the duration and the end of the reionization
phase.

\section*{Acknowledgments}
We thank R.~Salvaterra for clarifications about the CS07 results,
F.~Haardt, G.~Mellema, R.~Thomas and M.~Volonteri for useful
discussions, and the anonymous referee for several suggestions about
improving the manuscript. We also thank K. Visser for assistance in
working with the HPC cluster of the Centre for High Performance
Computing and Visualization of the University of Groningen, where most
of our computations were carried out. ER and MM thank the Institute
for Theoretical Physics of the University of Z\"urich, and the Kapteyn
Astronomical Institute of the University of Groningen for the
hospitality during the preparation of this paper. ER acknowledges
support from the Netherlands Organization for Scientific Research (NWO)
under project number 436016, and MM acknowledges support from the Swiss
National Science Foundation, project number 200020-109581/1
(Computational Cosmology \&{} Astrophysics).  SZ is a member of the
LOFAR project which is partially funded by the European Union, European
Regional Development Fund, and by ``Samenwerkingsverband
Noord-Nederland'', EZ/KOMPAS.



\end{document}